\documentclass[aip,amsmath,amssymb,reprint]{revtex4-1}

\pdfoutput=1
\usepackage{titlesec}

\usepackage[utf8]{inputenc}
\usepackage[english]{babel}
\usepackage[T1]{fontenc}
\usepackage{amsmath}
\usepackage{hyperref}
\usepackage{enumitem}

\usepackage{graphicx}
\usepackage{bm}
\usepackage{dsfont}
\usepackage{amsthm, amssymb,mathrsfs}
\usepackage{setspace}
\usepackage[toc,page]{appendix}
\usepackage{epstopdf}
\usepackage{mdframed}
\usepackage{verbatim} 
\usepackage{tcolorbox}

\newcommand{\beq}{\begin{equation}}
\newcommand{\beqs}{\begin{equation*}}
\newcommand{\eeq}{\end{equation}}
\newcommand{\eeqs}{\end{equation*}}
\newcommand{\beqa}{\begin{eqnarray}}
\newcommand{\eeqsa}{\end{eqnarray}}

\newcommand{\half}{\mbox{$\textstyle \frac{1}{2}$}}

\newcommand{\av}[1]{\langle #1\rangle}

\newcommand{\e}{\mathrm{e}}

\newcommand{\arFour}[4]{ \left( \begin{array}{cc} {#1} & {#2} \\ {#3} & {#4} \end{array} \right) }
\newcommand{\arFours}[4]{ \left( \arraycolsep=1pt\def\arraystretch{0.85} \begin{array}{cc} {#1} & {#2} \\ {#3} & {#4} \end{array} \right) }

\DeclareMathOperator{\sinc}{sinc}

\newcommand{\bqa}{\begin{eqnarray}}
\newcommand{\eqa}{\end{eqnarray}}

\newcommand{\radss}{\,rad$\cdot$s$^{-1}\,$}
\newcommand{\radsm}{\,rad$\cdot$m$^{-1}\,$}

\renewcommand{\thesubsubsection}{ \textbf{\Alph{subsubsection})}}
\titleformat{\subsubsection}[runin]
        {\normalsize\itshape}
        {\thesubsubsection}
        {0.5em}
        {}
        [ ---]
      
\definecolor{com}{rgb}{0.9,0.1,0.3}
\newcommand{\com}[1]{{#1}}
\definecolor{comtwo}{rgb}{0.95,0.1,0.0}
\newcommand{\comtwo}[1]{{#1}}

\begin{document}
\title{Backscattering in Nonlinear Microring Resonators Via A Gaussian Treatment of Coupled Cavity Modes}
\author{Will McCutcheon}
\affiliation{Quantum Engineering Technology Labs, H. H. Wills Physics Laboratory, University of Bristol, BS8 1FD, UK}
\affiliation{BBQLabs, Institute of Photonics and Quantum Sciences (IPAQS), Heriot-Watt University, Edinburgh, EH14 4AS, UK}
\email{w.mccutcheon@hw.ac.uk}

\date{\today}
\begin{abstract}
Systems of coupled cavity modes have the potential to provide bright quantum optical states of light in a highly versatile manner. Microring resonators for instance are highly scalable candidates for photon sources thanks to CMOS fabrication techniques, their small footprint \com{and the relative ease of coupling many such microrings together}, however, surface roughness of the wave-guides, and defects in the coupler geometry routinely induce splitting of the cavity modes due to backscattering and backcoupling. The parasitic back-propagating mode in the microring leads to hybridisation of the modes, altering the linear and nonlinear properties of this \com{system of coupled cavity modes}, and ultimately constraining the fidelity of quantum light sources that can be produced. In this paper, we derive a comprehensive general model for Gaussian nonlinear processes in \com{systems of coupled cavity modes}, based on an effective field Hamiltonian and a dispersive input-output model. The resulting dynamics of the equations of motion are evaluated in a Gaussian process formalism via the symplectic transformations on the optical modes. We \com{then use this framework to} numerically model and explore the problem of backscattering in microring resonators in physically relevant parameter regimes, involving the splitting of various resonances, we calculate the consequent impurity and heralding efficiency of various heralded photon schemes, we explore a perturbative explanation of the observations and assess the correspondence between spontaneous and stimulated processes in these systems. 
\end{abstract}

\maketitle

Microring resonators are a promising platform for delivering on-chip quantum light sources~\cite{Silverstone2016,Gentry2015,Harris2014}, offering high brightness~\cite{Savanier2016,Azzini2012,Jiang2015}, and increasingly high purities for interference~\cite{Faruque2018,Vernon2017}. However, even slight fabrication defects of waveguides leads to loss and backscattering~\cite{Morichetti2010,Morichetti2010b}, which in microrings causes coupling of the forward and backward propagating modes. \com{In the linear optical regime these effects have long been well understood~\cite{Little1997,Kippenberg2002}, are becoming well characterised experimentally~\cite{Morichetti2010a,Ballesteros2011}, including, asymmetric Fano splitting of the resonances~\cite{Li2016,Biasi2019} and methods to overcome these limitations have been demonstrated~\cite{Bogaerts2017,Li2019,Svela2020}. Microring cavities have great potential for uses in sensing applications~\cite{Yi2011,Yi2009,Yi2011a}, where backscattering often degrades performance, though  counterpropagating modes can be exploited for some sensing applications~\cite{Yang2018,DeGoede2021,Arbabi2011}}. Nonlinear optics remains predominantly constrained to single cavities~\cite{Garay-Palmett2013,Yang2007,Helt2010,Vernon2015}, with \com{multiple degenerate cavity modes} just considered in the continuous-wave (CW) pumping regime~\cite{Jiang2015}, alongside preliminary investigations into interrogation of the back-propagating modes~\cite{Suo2017}. Similarly, backscatter in cavities generating Kerr frequency combs often plays a significant role in degrading the quality of the resonances and subsequent phase stability~\cite{Fortier2019,Kippenberg2018,Chen2011}, though these classical models tend to focus on soliton formation in the steady states of the strongly nonlinear regime~\cite{Kondratiev2019,Kondratiev2020,Hill2020}, with counter-propagating solitons only recently being proposed~\cite{Yang2017} and demonstrated~\cite{Joshi2018a}. 

These systems however differ drastically from the \emph{Gaussian} regime in consideration here, in which we pursue properties of the spectral correlations between quantum fields of few cavity resonances in the presence of weak nonlinearities that can be well approximated by Hamiltonians \com{which are just quadratic in the quantum mode operators}, ie. \emph{Gaussian} nonlinearities. \com{This precludes application to Kerr Hamiltonians, which are quartic, were all the fields to be treated quantum mechanically, however in many instances some bright fields can be treated classically, resulting in quantum fields that evolve under Gaussian (quadratic) Hamiltonians.} Beyond backscattering in microrings, a general quantum mechanical treatment of the broadband spectral behaviors of a number of coupled \com{cavity modes}, coupled to a number of waveguides has wide application to the increasingly complex photonic devices being fabricated, from the linear dynamics of photonic atoms and molecules~\cite{Rakovich2010,Li2017,Zhang2019} through to next generation quantum light sources comprised of multiple cavities~\cite{Chuprina2018,Chuprina2019}. \comtwo{Whilst temporal coupled mode theory (TCMT) has been been demonstrated in the linear regime~\cite{Suh2004,Zhao2019a}, and for limited nonlinear systems~\cite{Chuprina2018,Heuck2019}, a general Hamiltonian based model for evaluating the Gaussian nonlinear quantum optical mode transformations induced by a several cavities, coupled to several waveguides, with arbitrary pumping and input states, remains to be demonstrated. Moreover, the Gaussian treatment here demonstrated exemplifies the nature of the linear symplectic inversion necessary to arrive at the full Gaussian solutions to such systems (applicable for arbitrary pumping and input fields), it invites the demonstration of a novel perturbative solution, as well as its polar decomposition, which provides insights into spontaneous and stimulated behaviors of general systems.}

\com{The result presented herein provides a system of equations that can be directly applied to a wide host of systems, mitigating the need to derive such equations on an \emph{ad hock} basis. Furthermore, in the pursued treatment different waveguide dispersion can be captured, and the solution to the system is the full Gaussian mode transformation~\cite{Braunstein2005,Olivares2012} easily facilitating the investigation of features beyond the often used first order approximations, such as applications in nonlinear interferometry~\cite{Chekhova2016}. In addition, a first order perturbative solution is presented, in the nonlinear interaction, that allows the linear dynamics of the system to be maintained to arbitrary order thereby helping to explain phenomena relating to the stimulated and spontaneous processes. The subsequent application to backscattering in photon pair generation, provides clear understanding of how the various effects manifest and how to accommodate them, which will directly aid in the design of a range of microring based devices.} 

To pursue a complete quantum optical model of such \com{systems of coupled cavity modes}, we draw closely upon the effective field methods developed to treat nonlinear optical processes in dispersive media in a canonical formalism~\cite{Sipe2004,Bhat2006,Sipe2009,Yang2008}. In particular, its applications to microring resonators~\cite{Vernon2015,Helt2010} which closely resembles input-output formalisms for TCMTs often used for such systems. This framework however is also applicable to modes of differing dispersion providing strict notions of mode orthogonality and normalisation~\cite{Sipe2009,Quesada2017}.

\section{The Model}

We consider a system of $J$ wave-guide `channels' with effective field annihilation operators $ \hat s_j(z,t) $, with $j=1,...,J$~\cite{Sipe2004,Yang2008}. The displacement field in these wave-guides is,
\begin{align*}
\hat{ \bm{D}}(\bm{r}) &= \sum_j \sqrt{\tfrac{\hbar \Omega_j}{2}} \tfrac{1}{\sqrt{2 \pi}} \e^{i k(\Omega_j) z}  \bm{d}^\perp_j (x,y) \hat s_j(z,t) + \text{h.c.}
\end{align*}
with transverse coordinates $x$ and $y$ and longitudinal coordinate $z$ being dependent on the field $j$.
The channel field $j$ is associated to the central frequency $\Omega_j$ and the dispersion relation $k_j(\Omega)$ for which we expand just to first order having group velocity $v_j$. 
\com{The normalisation prefactors ensure that the effective field operators obey the standard equal-time commutation relations ($[ \hat s_j(z,t),\hat s^\dagger_{j'}(z',t) ] =\delta_{jj'} \delta(z-z')$ ) providing the transverse field profiles are normalised accordingly (see Appendix~\ref{app:pumpPower}).}

There will exist multiple effective fields indexed by different $j$, occupying the same spatial mode over disjoint frequency intervals, resulting in effective fields which are \com{narrowband and thus slowly varying in space owing to the fast varying factor $\e^{i k(\Omega_j) z}$, which leads to additional terms arising in the EM Hamiltonian~\cite{Yang2008}. 
The different effective fields are associated to independent first order dispersion relations, meaning that higher-order dispersive properties of the waveguides may be present in the model, however the individual narrowband effective fields are subject to only linear dispersion. As such, we require that each effective field is sufficiently narrowband to neglect higher order dispersion.}

These are coupled at $z=0$ to $N$ cavities modes with annihilation operators $\hat a_n(t)$ with $n=1,...,N$, and energy $\omega_n$. The coupling between channel mode \com{$j$} and cavity mode \com{$n$} is given by the channel-cavity coupling elements $\gamma_{nj}$. The cavity mode $n$ is coupled to cavity mode $m$ with cavity-cavity coupling $g_{nm}$. Finally channel mode $j$ is coupled to channel $k$ (at $z=0$) by channel-channel couplings $C_{jk}$. We note that such a point coupling model is only possible when group velocities are matched so insist that $C_{jk} \neq 0 \Rightarrow v_j=v_k$. The complete linear Hamiltonian is
\begin{widetext}
\beq
\begin{aligned}
\label{eqs:LinearHamiltonian}
H^{(L)} &= \hbar\sum_n  \omega_n \hat a^\dagger_n \hat a_n  + \hbar\sum_j \biggl(  \Omega_j \int d z \hat s_j^\dagger (z) \hat s_j (z) + \frac{ i  v_j}{2} \int dz \bigl( \frac{d  \hat s_j^\dagger(z)}{dz}\hat s_j(z)-  \hat s_j^\dagger (z)\frac{d \hat s_j(z)}{dz}\bigr)\biggr) \\
&\quad + \hbar\sum_{nl}  \gamma_{nl}  \bigl( \hat a_n^\dagger \hat s_{l} (0)+ h.c. \bigr) \,  + \hbar \sum_{nm}g_{nm}\hat a^\dagger_n \hat a_m  \,  + \hbar\sum_{jl}  C_{jl}  \hat s^\dagger_{j} (0) \hat s_{l} (0) \, .
\end{aligned}
\eeq
\end{widetext}
\com{The first term is the free Hamiltonian for the cavity modes and the third, four and fifth terms describe channel-cavity, cavity-cavity and channel-channel coupling respectively. The second term describes the free Hamiltonian of the channels, exhibiting terms arising from the series (linear) expansion of the dispersion relation and a subsequent partial integration~\cite{Sipe2004,Quesada2020}.} 
 Let $V$, $\Omega$ and $\omega$ be diagonal matrices of group velocities ($v_i$), carrier frequencies of effective fields ($\Omega_i$) and cavity modes ($\omega_i$) respectively. And we denote the matrices $C$, $g$ and $\gamma$ having elements $ C_{jl}$, $g_{nm}$ and $\gamma_{nj}$ respectively so that $g=g^\dagger$ and $C=C^\dagger$.

We focus on the nonlinear Hamiltonian resulting from $\chi^{(3)}$ response of the material which will be assumed to only occur in the cavities, in which the field amplitudes are large and may be described by,
\beq
\begin{aligned}
\label{eqs:NonlinearHamiltonianRaw}
H^{(NL)} &=\hbar \sum_{nn'mm'}\lambda_{nn'mm'} \bigl( \hat a^\dagger_n \hat a^\dagger_{n'} \hat a_{m} \hat a_{m'} + \text{h.c.} \bigr) \, ,
\end{aligned}
\eeq
with $\lambda$ to be determined by the underlying cavity geometry and material properties (see Appendix~\ref{app:NLLambda}).
Similarly, $\chi^{(2)}$ processes could be tackled in a more straightforward manner. \com{We will neglect nonlinear losses, such as two-photon- and free-carrier- absorption~\cite{Rosenfeld2019}}.   

\subsection*{Linear Equations of Motion}
We construct vectors containing effective field operators in the channel and cavity and denote them by bold operators, $\bm s(z,t) = \bigl(\hat s_1(z,t),\hat s_2(z,t),...,\hat s_J(z,t)\bigr)^\top$ and $\bm a(t) = \bigl(\hat a_1(t),\hat a_2(t),...,\hat a_N(t)\bigr)^\top$. The channel mode operators obey the Heisenberg equations of motion (EOMs),
\beq
\begin{aligned}
\label{eqs:LinearChannelEOM}
(\frac{\partial }{ \partial t}+ V \frac{\partial}{\partial z} +i \Omega ) \bm s(z,t) =-i \gamma^\dagger \bm a(t)\delta(z) - i C \bm s(0,t) \, .
\end{aligned}
\eeq
The fields entering and exiting the coupling region are given by the one-sided limits $\bm s^\pm(0,t):= \lim_{z\rightarrow 0^\pm} \bm s(z,t)$ and at the discontinuity we impose $\bm s(0,t)=\half(\bm s^+(0,t)+\bm s^-(0,t))$. We may then compactly express the input-output relations (see Appendix~\ref{app:inputOutput}),
\begin{align}
\bm s^+(0,t)&=T\biggl( \bm s^-(0,t) - i  V^{-1} \bar \gamma^{\dagger} \bm a(t)\biggr) \\
\text{with} \quad  \tilde C &:=(\mathds{1} + \tfrac{i}{2}V^{-1} C) \nonumber \\
T&:=\tilde C^{-1}\tilde C^\dagger \quad,\quad \bar \gamma:=\gamma \tilde C^{-1} \nonumber
\end{align}
Note that $T:=\tilde C^{-1}\tilde C^\dagger= \exp\{-2 i \tan^{-1}(V^{-1}C/2)\}$ is explicitly unitary and can asymptotically approach arbitrary passive transformations. 
We first consider the linear equations of motion ($H^{(NL)}=0$) for the cavity \com{modes},
\beq
\begin{aligned}
\label{eqs:RawCavityEOM}
 \biggl(\frac{d}{dt}  + i \omega + \half \bar \gamma  V^{-1} \gamma^\dagger +i  g \biggr) \bm a (t)&=- i \bar \gamma  \bm  s^- (0,t) \, ,
\end{aligned}
\eeq
which, in terms of the slowly varying envelope operators $\tilde {\bm a}(t) = \exp(i  \omega t) \bm a(t)$ and $\tilde {\bm s}(0,t) = \exp(i  \Omega t) \bm s(0,t)$\footnote{Whilst this rotating frame transforms the operators to slowly varying operators, no approximation is being made here. In fact the slow variation of the fields is implied by the effective field expansion whereby a narrow bandwidth is required in order to expand the dispersion relations to first order.}, read,
\begin{align}
\label{eqs:linearEOMsTime}
\biggl(\frac{d}{dt}   +\bar \Gamma(t) \biggr) \tilde {\bm a} (t)=&- i \bar \gamma(t)  \tilde {\bm s}^- (0,t)\\
\text{with}\quad \quad \bar \Gamma:=&\half \bar \gamma  V^{-1} \gamma^\dagger +i  g \nonumber\\
\bar \Gamma(t):=\e^{i  \omega t}\bar \Gamma\e^{-i  \omega t}&, \quad
\bar \gamma(t) :=\e^{i  \omega t}\bar \gamma \e^{-i  \Omega t}\nonumber
\end{align}
Linear transformations admit frequency domain solutions in terms of the Fourier transformed operators,
\beqs
\begin{aligned}
 \bm{\tilde s}^\pm(z,t) &= \int  \tfrac{d k}{\sqrt{2\pi}} \mathbf { \tilde s}^\pm(k)\e^{i k z} \e^{- i V k t}\\
\end{aligned}
\eeqs
To facilitate moving to a frequency domain model we introduce the phenomenological group velocities, $\mathcal{V}$, for the cavity modes. Then the Fourier amplitudes for the cavity fields are defined 
\beqs
\begin{aligned}
\tilde{\mathbf{a}}(k)= \mathcal{V} \int  \tfrac{d t}{\sqrt{2\pi}} \tilde{\bm{a}}(t) \e^{i \mathcal{V} k t} \\
\end{aligned}
\eeqs
Then by induction we find \com{(see Appendix~\ref{app:FourierInduction})},
\beq
\begin{aligned}
\label{eqs:FourierLinear}
\int  d k'   \biggl(-i \mathcal{V} k' \delta(k-k')   + \bar \Gamma(k,k') \biggr)  \tilde {\mathbf a} (k')\\
=-i  \int   d k'  \bar \gamma(k,k')      {\mathbf s}^- (k') \\
\end{aligned}
\eeq
where we have defined,
\beqs
\begin{aligned}
\bar \Gamma(k,k')&:= 
\mathcal{V} \int  \tfrac{d t}{2\pi}\e^{i (\mathcal{V} k +  \omega ) t}\bar \Gamma\e^{-i(\mathcal{V} k'+  \omega) t} \\
 \bar \gamma(k,k')&:=
 \mathcal{V} \int  \tfrac{d t}{2\pi}\e^{i (\mathcal{V} k + \omega ) t}\bar \gamma \e^{-i({V} k'+  \Omega) t}\\
\end{aligned}
\eeqs
This linear system can be solved exactly in terms of the absolute frequency as shown in Appendix~\ref{app:LinearAbsoluteFrequency}. We often consider however, the case of equal tunings and group velocities everywhere $\bar \Gamma$ and $\bar \gamma$ have support, ie. $\gamma_{nj}\neq 0 \Rightarrow V_{jj}=\mathcal{V}_{nn}$ and $g_{nm}\neq 0 \Rightarrow \omega_n=\omega_m$. In this case both integrands become proportional to $\delta(k-k')$ and we have,
\beq
\begin{aligned}
\label{eqs:TunedLinearFrequencyDomainSol}
\tilde {\mathbf a} (k)&=- i\biggl(-ik\mathcal{V}  + \bar \Gamma \biggr)^{-1} \bar \gamma  \tilde {\mathbf s}^- (k) \, .
\end{aligned}
\eeq
For which the transmitted fields become,
\beq
\label{eqs:linearInputOutput}
\tilde{\mathbf s}^+(k)
= T \biggl(\mathds{1} -  V^{-1} { \bar \gamma}^\dagger\bigl(-ik \mathcal{V} + \bar \Gamma \bigr)^{-1} \bar{ \gamma}\biggr)  \tilde {\mathbf s}^- (k) \,.
\eeq
These solutions to the linear equations of motion constitute a system of equations equivalent to those encountered in existing literature on the temporal coupled mode theory of multiple optical cavities. This approach however, derived from an underlying Hamiltonian which is explicitly Hermitian, avoids redundancies in parameterisation~\cite{Fan2003,Suh2004}. The precise relationship between TCMT models and the more general quasi-normal modes of such systems we leave for future work~\cite{Lang1988,Zhao2019a}.

\subsection*{Nonlinear Equations of Motion}
With the inclusion of the nonlinear Hamiltonian (Eqs.~\ref{eqs:NonlinearHamiltonianRaw}), we obtain additional terms in the EOMs for the cavity modes, Eqs.~\ref{eqs:RawCavityEOM}. In particular, to the $\hat a_n(t)$ element on the l.h.s.~of Eqs.~\ref{eqs:RawCavityEOM}, we must introduce the terms
$ i\sum_{n'mm'}\lambda_{nn'mm'} \hat a^\dagger_{n'} \hat a_{m} \hat a_{m'}$. When all fields are treated quantum mechanically closed form solutions to the EOMs can rarely be found, and for bright quantum fields numeric approaches do not yield well to perturbative methods. At this stage, we move to the Gaussian regime and declare which fields are \emph{signal} fields which will be assumed relatively weak and treated quantum mechanically. These include both \emph{signal} and \emph{idler} fields from conventional treatments of four-wave mixing (FWM), though owing to the multiplicity or degeneracy of these fields they are all considered the \emph{signal} fields. In contrast the \emph{pump} fields are those bright fields which may be approximated by their classical mean fields which will obey the same equations of motion, with the operators replaced with the mean fields $\av{\hat{a}_{p_n}(t)}$. These bright \emph{pump} fields obey cubic equations of motion when including all nonlinear effects, though they can be tractably numerically solved independently providing we neglect the negligible effects arising from back-action from the weak \emph{signal} fields, ie. we exclude nonlinear coupling to the quantum \emph{signal} modes in the pump modes' EOMs. This approximation effectively neglects any annihilation of \emph{pump} photons through coupling to signal fields and is called the \emph{undepleted pump approximation}. In addition, nonlinear terms involving only signal fields can be safely neglected from their EOMs. \com{This regime prevents the (quartic) Kerr Hamiltonian from acting on the quantum fields directly, and instead is capable only of dealing with instances where two (of the four) fields it acts on need to be treated quantum mechanically, resulting in, at most, quadratic terms in the quantum EOMs.}

We have then two systems of equations: 1) the classical EOMs for the pump fields, 2) the quantum mechanical \com{quadratic} EOMs for the signal fields driven by time-dependent contributions from the pump fields. These signal fields then exhibit time-dependent quadratic nonlinear interactions, hence their Gaussian nature, and these come in two forms. \emph{Phase modulation} - which contributes terms of the form $i\sum_{pn'p'}\lambda_{npn'p'}\av{\hat a^\dagger_{p}} \av{\hat a_{p'}} \hat a_{n'}$ which resembles a time varying linear process and includes frequency conversion when $n'\neq n$. And \emph{Squeezing} - in which the evolution of the creation and annihilation operators becomes coupled together via terms of the form  $i\sum_{n'pp'}\lambda_{nn'pp'}\av{\hat a_{p}} \av{\hat a_{p'}} \hat a^\dagger_{n'}$. 
Note that even in the case of perturbative methods, such as photon pair generation considered below, it is implicit that the system is in the Gaussian regime in order to obtain the quadratic Hamiltonian for declared quantum fields.

To accommodate these quadratic terms in the EOMs, it is convenient to adopt the formalisms of Gaussian quantum optics\com{~\cite{Braunstein2005,Olivares2012}} and introduce the capitalised vectors of operators containing both creation and annihilation operators,  $\tilde{ \bm{A}} (t) = (\tilde{\bm{a}}^\top (t),\tilde{\bm{a}}^{ \dagger} (t))^\top$ and $\tilde{ \bm{S}}^\pm (0,t) = (\tilde{\bm{s}}^{\pm T} (0,t),\tilde{\bm{s}}^{\pm \dagger} (0,t))^\top$. By doing so we may express the full nonlinear equations of motion for the signal fields in the cavity,
\begin{align}
\label{eqs:nonlinearEOMsTime}
\biggl(\frac{d}{dt}   +&\bar {\bm{\Gamma}}(t) \biggr) \tilde {\bm A} (t)=- i \bar {\bm{\gamma}}(t)  \tilde {\bm S}^- (0,t) ,\\
\bar {\boldsymbol{\Gamma}}(t)&:=\left( \begin{array}{cc} {\bar \Gamma (t) +\bar \Gamma^{(\text{PM})}(t)}&{\bar \Gamma^{(\text{Sq})}(t)}\\
{\bar \Gamma^{(\text{Sq})*}(t)}&{\bar \Gamma^*(t)+\bar \Gamma^{(\text{PM})*}(t)} \end{array} \right)\nonumber,\\
\bar {\bm{\gamma}}(t) &:=\left( \begin{array}{cc} {\bar \gamma (t)}&{0}\\
{0}&{\bar \gamma^*(t)} \end{array} \right),\nonumber
\end{align}
where the elements of the nonlinear parts are
\beq
\hspace{-0.02\columnwidth}
\begin{aligned}
(\bar \Gamma^{(\text{Sq})}(t))_{s i}&=i \sum_{pp'}  \lambda_{s i p p' } \av{\tilde a_{p}(t)}\av{\tilde a_{p'}(t)}\e^{i(\omega_s + \omega_i -\omega_p-\omega_{p'}) t},\\
(\bar \Gamma^{(\text{PM})}(t))_{s i} &= i \sum_{pp'}  \lambda_{s p i p' } \av{\tilde a_{p}(t)}\av{\tilde a^\dagger_{p'}(t)}\e^{i(\omega_s - \omega_i +\omega_p-\omega_{p'}) t}.
\label{eqs:nonlinearGammaTerms}
\end{aligned}
\eeq

Solutions to these EOMs are in fact linear symplectic transformations - the central feature of Gaussian quantum optics. We will from hereon use bold uppercase symbols to denote these kinds of vectors of operators and the symplectic operations which act on them. Whilst one could proceed by propagating the system under an appropriate Greens' function, in the absence of phase modulation, it is more straightforward to work again in the frequency domain.

In particular, we pursue a frequency domain Gaussian input-output relation of the form,
\beqs
\begin{aligned}
\tilde{\mathbf S}^+(k)&=\int dk'  \mathbf{M}(k,k') \tilde{\mathbf S}^{-}(k'),
\end{aligned}
\eeqs
with the linear symplectic operator $\mathbf{M}(k,k')$ to be defined below. It is convenient to define,
\beqs
\begin{aligned}
\bm{{V}}:=&\arFours{{V}}{0}{0}{{V}} \, , \, \bm{\mathcal{V}}:=\arFours{\mathcal{V}}{0}{0}{\mathcal{V}}\,,\,
\bm{\mathcal{I}}^- :=\arFours{\mathds{1}}{0}{0}{-\mathds{1}},\\
\bm{\mathcal{V}}^- =&\bm{\mathcal{I}}^- \bm{\mathcal{V}}\quad , \qquad
\bm{{V}}^- =\bm{\mathcal{I}}^- \bm{{V}}\,,
\end{aligned}
\eeqs
so that the frequency domain Fourier transforms are
\beqs
\begin{aligned}
\mathbf {\tilde S}^\pm(k)&= \bm{{V}} \int  \tfrac{d t}{\sqrt{2\pi}} \tilde{\bm S}^\pm(0,t) \e^{i \bm{{V}}^- k t}, \\
\tilde{\mathbf{A}}(k)&= \bm{\mathcal{V}} \int  \tfrac{d t}{\sqrt{2\pi}} \tilde{\bm{A}}(t) \e^{i \bm{\mathcal{V}}^- k t}. \\
\end{aligned}
\eeqs
Inline with Eqs.\ref{eqs:FourierLinear}, by induction we derive \com{(see Appendix~\ref{app:FourierInduction})},
\begin{align}
\int  d k'   \biggl(-i \bm{\mathcal{V}}^- k' \delta(k-k')   + \bm{\bar \Gamma}(k,k') \biggr)  \tilde {\mathbf A} (k'),\nonumber\\
=-i  \int   d k'  \bm{\bar \gamma}(k,k')      {\mathbf S}^- (k'),
\label{eqs:FourierNonLinear}
\end{align}
where 
\beq
\begin{aligned}
\label{eqs:bigGammaOfK}
\bm{\bar \Gamma}(k,k')&:= \bm{\mathcal{V}} \int  \tfrac{d t}{2\pi}\e^{i \bm{\mathcal{V}}^- k t}\bm{\bar \Gamma}(t)\e^{-i \bm{\mathcal{V}}^- k' t}, \\
\end{aligned}
\eeq
For some cavity modes indexed $p_r$ and $p_{r}'$, containing classical pump fields $\av{\tilde a_{p_r}(t)}=\int \left(dk/\sqrt{2\pi}\right) \av{\tilde{\mathrm{a}}_{p_r}(k)} \e^{-i v_{p_r} k t}$ (similarly for $p_{r}'$) and approximately energy matched signal and idler mode $s_r$ and $i_r$ so that $\lambda_{p_r p_{r}' s_r i_r }$ is non-negligible, the corresponding component can be evaluated
\begin{align}
\label{eqs:nonlinearConvolution}
 \biggl(\mathcal{V} \int  \tfrac{d t}{2\pi} 
&\e^{i \mathcal{V} k t}\bar \Gamma^{(FWM)}(t)\e^{i \mathcal{V} k' t}\biggr)_{s_r i_r} \nonumber \\
= &i v_{s_r} \sum_{p_r p_{r'}}\lambda_{p_r p_{r}' s_r i_r } \tfrac{1}{2\pi} 
\tfrac{v_{p_{r}}}{v_{p_{r}'}} \mathcal{C}_{p_r p_r' s_r i_r},\\
\mathcal{C}_{p_r p_r' s_r i_r}:=&\av{\tilde{\mathrm{a}}_{p_{r}}}*\av{\tilde{\mathrm{a}}_{p_{r}'}}(\frac{ v_{s_r} k +v_{i_r} k' + \Delta}{v_{p_{r}'}}),\nonumber 
\end{align}
where $\left(\av{\tilde{\mathrm{a}}_{p_r}}*\av{\tilde{\mathrm{a}}_{p_{r}'}}\right)(\cdot)$ is the convolution. 
All that remains to be done, is inversion of the scalar valued linear transformation on the $l.h.s.$ of Eqs.~\ref{eqs:FourierNonLinear}, resulting in an explicit solution for $\tilde {\mathbf A} (k)$, followed by a substitution into the input-output relation
\beq
\begin{aligned}
\label{eqs:fullSolution}
\tilde{\mathbf{S}}^+(k)&=  \bm{T} \bigl( \tilde{\mathbf {S}}^-(k) - i \bm{V}^{-1} \bm{\bar \gamma}^\dagger \tilde {\mathbf A}(k)\bigr) \\
&=\int dk'  \mathbf{M}(k,k') \tilde{\mathbf S}^{-}(k').
\end{aligned}
\eeq
This function, $\mathbf{M}(k,k')$ entirely characterises the linear and nonlinear response of system. It is, by construction, explicitly symplectic and serves as the focus of the forthcoming discussion. On the existence of closed form analytic solutions to this system, we observe that closed form inverses to the function valued matrix in Eqs.~\ref{eqs:FourierNonLinear} limit the practicality of such expressions in all but some particularly convenient examples. Insight can be gleaned by considering the Neumann series expansion of this inverse about $\Gamma^{(\text{Sq})}$, which displays the characteristic relationship between Gaussian transformations and their underlying Hamiltonian, in particular, elements of $\mathbf{M}(k,k')$ contain hyperbolic trigonometric functions of $\Gamma^{(\text{Sq})}$ (along with a linear transformation). For small nonlinearities, perturbative methods can be obtained by truncating this series to first order (see appendix~\ref{app:pertubative}), which amounts to the approximation $\text{sinh}(\Gamma^{(\text{Sq})}) \approx \Gamma^{(\text{Sq})}$. We add, that in contrast to many perturbative methods, this approximation can be applied whilst maintaining the full linear response of the system, which is integral for applications involving non-trivial input states to the transformation(to be discussed in Section~\ref{sec:SET}).

\com{We provide a summary of how this framework is applied.

\begin{tcolorbox}
\section*{Application of the Framework}
\com{
\begin{enumerate}
    \item Identify the pump fields' central frequencies, group velocities and couplings $\Omega_j,V_j,\omega_j,\mathcal{V}_{j},\gamma_{ij}, g_{ij}, C_{ij}$ and the input pump fields $\tilde {\mathbf s}^{-}_j (k)$ for $i,j$ indexing \emph{pump} fields.
    \item Construct Eqs.~\ref{eqs:FourierLinear} and solve for the pump cavity fields $\tilde {\mathbf a}_j (k)$ with $j$ indexing \emph{pump} fields, (possibly via Eqs.~\ref{eqs:TunedLinearFrequencyDomainSol} or Eqs.~\ref{eqs:TunedLinearFrequencyDomainInputOutputSol})
    \item Identify the signal fields' parameters: $\Omega_j,V_j,\omega_j,\mathcal{V}_{j},\gamma_{ij}, g_{ij}, C_{ij}$ for $i,j$ indexing \emph{signal} fields..
    \item Identify the relevant nonzero nonlinear couplings $\{\lambda_{s_1 s_2 p_1 p_2}\}$ (via Appendix~\ref{app:NLLambda})
    \item Construct $\bm{\bar \Gamma}(k,k')$ as per Eqs.~\ref{eqs:bigGammaOfK} using Eqs.~\ref{eqs:nonlinearGammaTerms} (see for instance Eqs.~\ref{eqs:nonlinearConvolution})
    \item Construct Eqs.~\ref{eqs:FourierNonLinear}, to obtain $\tilde {\mathbf A}(k)$ and substitute into Eqs.~\ref{eqs:fullSolution} to obtain the full transformation $\mathbf{M}(k,k')$. 
    \item[\refstepcounter{enumi}(*\number\value{enumi})] \comtwo{Evaluate Eqs.~\ref{eqs:perturbativeInputOutput}, for perturbative input-output solution.}
\end{enumerate}
}
\end{tcolorbox}

We see that parts 1) and 3) trivially involve gathering the relevant information describing the system being modelled. Part 2) involves solving the dynamics of the pump fields in the cavity modes. Since the pump fields are scalar, this could be achieved by numerically solving Eqs.~\ref{eqs:nonlinearEOMsTime} with the inclusion of phase modulation. Part 4) consists of identifying the strength of the nonlinear interactions between the various pump and signal fields. Parts 5) and 6) then result in the signal fields' Gaussian transformation due to the cavity modes and the nonlinear interaction with the pump fields. 

A wide variety of systems, defined by the parameters outlined in parts 1),3) and 4), can be readily solved in this framework. One such example from Chuprina et al~\cite{Chuprina2019,Chuprina2018} is given in Appendix~\ref{app:Applications}. We turn for now to another application, backscattering in photon pair sources, which despite having widespread ramifications on real world devices, has yet to be presented. 
}

\section{Backscattering in microring Resonators}
Surface roughness of wave-guides in integrated platforms is routinely a cause of propagation losses. Whilst some of this scattering is lost to bulk/environment modes, some of the field is scattered into backward-propagating modes in the wave-guides. In microring resonators, this counter-propagating mode constitutes an additional cavity \com{mode}, and the scattering causes coupling between them. Coupling of cavities induces hybridisation and splitting of their resonance, and this can cause degradation of the quantum light sources they aim to generate. We consider the effects of backscattering in a microring resonator designed to achieve degenerately pumped four-wave mixing. We first solve the linear equations of motion for the pump field in the frequency domain, and use this to drive the nonlinear effects on the signal and idler fields. 

The bus wave-guide supports forward and backward propagating effective fields at the pump frequency $\Omega_p$ with group velocities $v_p$, denoted $\hat s_{p1f}(z,t)$ and $\hat s_{p1b}(z,t)$. An additional `phantom' channel is introduced to induce loss in the ring and has effective pump fields $\hat s_{p2f}(z,t)$ and $\hat s_{p2b}(z,t)$, chosen (for convenience) to have equal carrier frequency and group velocity. The microring supports an anti-clockwise (forwards) and clockwise (backwards)  mode at pump frequency ($\omega_p=\Omega_p$) denoted by $\hat a_{pf}(t)$ and $\hat a_{pb}(t)$ respectively. In total the mode operators we need to consider are $\bm a_p(t) = \bigl(\hat a_{pf}(t),\hat a_{pb}(t)\bigr)^\top$  and $\bm s_p (t) = \bigl(\hat s_{p1f}(z,t),\hat s_{p1b}(z,t),\hat s_{p2f}(z,t),\hat s_{p2b}(z,t)
\bigr)^\top$. We consider a ring designed to be critically coupled with coupling parameter $\gamma_p$, perturbed by backscattering in the ring characterised by $g_p$, backcouplings (channel-cavity coupling) $\delta_{fb}$ and $\delta_{bf}$ (taken proportionally to $\gamma_p$), and channel backreflection $c_p$, so that our Hamiltonian is determined by the matrices,
\beq
\begin{aligned}
\label{eqs:BackscatterParamsPump}
{\gamma} &= \gamma_p \left( 
\begin{array}{cccc} {1 }&{\delta_{fb}}&{1}&{0}\\
{\delta_{bf}}&{1}&{0}&{1} \end{array} \right), \,\, 
{g} =  \left(
\begin{array}{cc} {0 }&{g_p}\\
{g_p^*}&{0} \end{array} \right)  , \\
{C} &=  v_p\left(
\begin{array}{cccc} {0 }&{c_p }&{0 }&{0 }\\
{c_p^* }&{0 }&{0 }&{0 }\\
{0 }&{0 }&{0 }&{0 }\\
{0 }&{0 }&{0 }&{0 }\end{array} \right)\,.\\
\end{aligned}
\eeq
From Eqs.~\ref{eqs:TunedLinearFrequencyDomainSol} and Eqs.~\ref{eqs:linearInputOutput} we can promptly obtain the pump field in the ring and the resulting input-output relations. 

We consider Silicon-on-insulator wave-guides with cross section $220$nm-by-$500$nm pumped close to $1.550 \mu \text{m}$ for which the effective index can be found by simulation to be $n_{eff}(\lambda)=2.44 - 1.13(\lambda - 1.55) - 0.04(\lambda - 1.55)^2$. The pump will have central wavelength $1.546 \mu \text{m}$ (channel 39 of the C-Band ITU grid) so that \mbox{$\Omega_p = 1.2183 \times 10^{15}$ \com{\radss}} and $v_p=7.1532\times 10^7\text{ms}^{-1}$. The microring will have circumference $202.56\mu \text{m}$, and an amplitude round trip transmission (self-coupling) $\sigma_r=0.985$, which corresponds to $\gamma_p=8.7277\times 10^8$ \com{\radss} $= \sqrt{\frac{2 (1-\sigma_r ) v_p^2}{L_r}}$. We will take as input pump field a top-hat in frequency across the simulation range, which would in practice be well approximated by a sufficiently broadband pulse filtered through a dense wavelength division multiplexer, and permits us to focus on the properties of the system, though we note that more elaborate pumping schemes could be used to tailor the generated states further~\cite{Christensen2018}. \com{Since the power coupled to the ring is dominated only by the overlap with the ring resonance, whereas the extent of the tophat remains a simulation feature, the pump power is not a meaningful quantity in this case. In practice some finite bandwidth pump pulse would be chosen, so, for reference, an example with narrowband pumping is included in Appendix~\ref{app:pumpPower}. } 

In Fig.~\ref{fig:pumpSplittingLinear}, we plot the transmitted fields, intra-cavity fields, and their convolutions. The transmitted fields to the bus modes indicate features one could access by interrogation of the ring under linear excitation, whereas the intra-cavity powers, which would in general require fitting of the under-laying linear parameters, indicate toward the resonant enhancement and escape probability of generated photons. The convolution of the intra-cavity power of the pump field informs us of the nonlinear contribution to Eqs.~\ref{eqs:nonlinearConvolution}, and together with the forthcoming perturbative treatment, these features dictate the generated states. In part a) we \com{choose a set of parameters which reproduce the routinely observed asymmetric splitting of the transmitted field resonance, as well as slight coupler backreflection}, $g_p=10^{10}$\com{\radss}, $\delta_{fb}=0.2$, $\delta_{bf}=0$ and $c=0.2$\com{\radsm}. We note that these fields are dependent on the phase of the coupling parameters, for instance, when $\arg(\delta_{fb})=\pi/2$, the field becomes symmetric about $k=0$. With a solution for the pump dynamics, we are free to explore how these generate nonlinear effects on the signal and idler fields. Whilst full knowledge of a systems linear response would allow the nonlinear properties to be evaluated, we will focus on few parameter models, for which parts \emph{i)} through \emph{iv)} indicate the various fields for $g_p=\{1/5, 8/5, 22/5, 49/5\}*10^{10}$\com{\radss} (with other parameters $0$).  
\begin{figure}[ht]
\centering
\vspace*{0in}
\hspace*{-0.02\columnwidth}
\includegraphics[width=1.04\columnwidth]{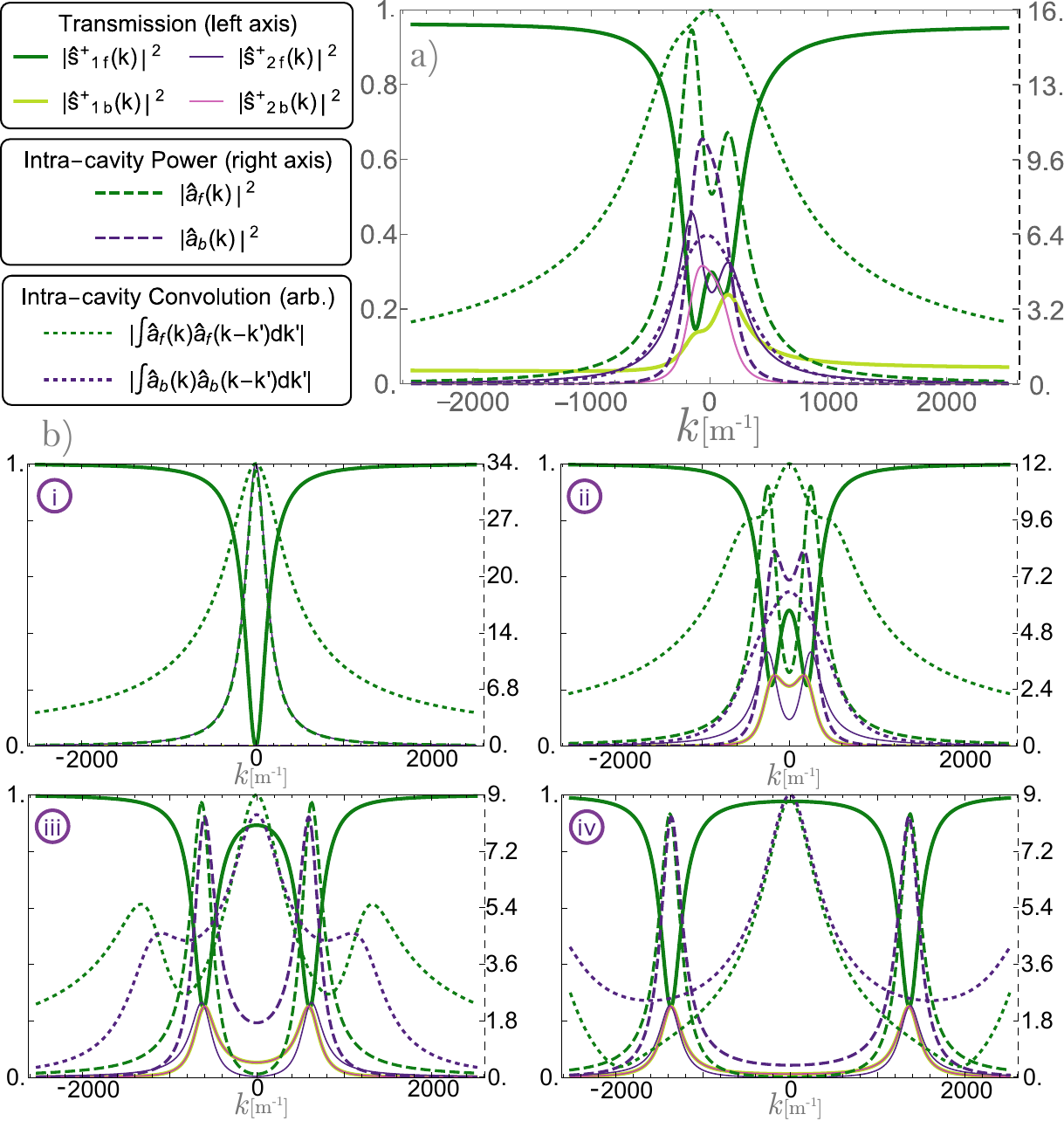}
\caption{ Transmitted fields, intra-cavity fields, and auto-convolution of intra-cavity fields for various backscattering parameters. Part a) $g_p=10^{10}$\com{\radss}, $\delta_{fb}=0.2$, $\delta_{bf}=0$ and $c_p=0.2$\com{\radsm}. 
Parts \com{b)} \emph{i)}, \emph{ii)}, \emph{iii)} and \emph{iv)} show simple resonance splitting by parameter $g_p=\{1/5, 8/5, 22/5, 49/5\}*10^{10}$\com{\radss} respectively. Whilst the intra-cavity field continues to split as $g_p$ increases, the auto-convolution of the field exhibits three peaks, with the middle of these coming to dominate.}
\label{fig:pumpSplittingLinear}
\end{figure}

\com{For the purpose of the present manuscript, exploring photon pair generation, where low pump powers can be used at the expense of low generation rates, we will neglect self-phase modulation of the pump, as well as cross-phase modulation, and focus solely on how backscattering affects the generated photons through four-wave mixing.} This nonlinear effect only become appreciable close to energy- and phase- matched regions, so we will consider just instances in which some signal and idler cavity fields obey $2\Omega_{p}\approx\Omega_{s}+\Omega_{i}$, and we will include both clockwise and counter-clockwise fields as before. For each of the disjoint spectral intervals, effective fields in the bus and 'phantom' channels are introduced, both forward and backward propagating so our \emph{signal} fields consist of $\bm a(t) = \bigl(\hat a_{sf}(t),\hat a_{sb}(t),\hat a_{if}(t),\hat a_{ib}(t)\bigr)^\top$  and $\bm s (t) = \bigl(\hat s_{s1f}(z,t),\hat s_{s1b}(z,t),\hat s_{s2f}(z,t),\hat s_{s2b}(z,t),\hat s_{i1f}(z,t),\allowbreak \hat s_{i1b}(z,t),\hat s_{i2f}(z,t),\hat s_{i2b}(z,t)\bigr)^\top$, so the indexes $s$ ($i$) relate to signal (idler) spectral intervals, $1$ and $2$ to the bus channel and phantom channel, and $f$ and $b$ to the forward and backward propagating fields respectively.
We consider signal and idler resonance to be next-but-one neighbours of the pump across the frequency comb of the microring, so that $\Omega_s =1.2208 \times 10^{15} \text{\radss}$ ($1.543 \mu \text{m}$ ITU channel 43) and $\Omega_i = 1.2157 \times 10^{15} \text{\radss}$ ($1.549 \mu \text{m}$ ITU channel 35), $v_s=7.1538\times 10^7 \text{ms}^{-1}$ and $v_i=7.1525\times 10^7 \text{ms}^{-1}$. Similarly to the pump fields, we introduce a complete set of parameters of the form Eqs.~\ref{eqs:BackscatterParamsPump} for both signal and idler.
We will take self-coupling of the ring to be constant so that $\gamma_s=\gamma_i=\gamma_p$, ie. the length of the coupling region is sufficiently long that the coupling is achromatic across the region of interest. \com{This achromatic coupling is know to cause impurity in the generated photons even in the absence of backscattering, and whilst methods to overcome these limitations have been proposed\cite{Vernon2017}, this coupling well approximates the most simple design; a microring evanescently coupled to a waveguide.}

The complete nonlinear contributions are,
\beq
\label{eqs:nonlinearBackscatterTerm}
\begin{aligned}
& \biggl(\mathcal{V} \int  \tfrac{d t}{2\pi} 
\e^{i \mathcal{V} k t}\bar \Gamma^{(FWM)}(t)\e^{i \mathcal{V} k' t}\biggr)\\
&= i \tfrac{1}{2\pi}  \lambda \mathcal{V}
{\small
\medmuskip=-2mu
\thinmuskip=-2mu
\thickmuskip=-2mu
\nulldelimiterspace=-1pt
\scriptspace=0pt
\left( \begin{array}{cccc} {0}&{0}&{\mathcal{C}_{p_f p_f s_f i_f}}&{0}\\
{0}&{0}&{0}&{\mathcal{C}_{p_b p_b s_b i_b}}\\
{\mathcal{C}_{p_f p_f i_f s_f}}&{0}&{0}&{0}\\
{0}&{\mathcal{C}_{p_b p_b i_b s_b}}&{0}&{0}\end{array} \right)}
\end{aligned}
\eeq
We have then a complete model to explore the consequences of backscatter and backcoupling in a physically relevant setting. We will numerically solve Eqs.~\ref{eqs:fullSolution}, and discuss the features in the symplectic tensor $\mathbf{M}(k,k')$. 
To understand what these transformations imply for photon pair generation in the spontaneous regime, we first perform a polar decomposition, so that $\mathbf{M}(k,k') = \mathbf{M}^h(k,k')\mathbf{M}^u(k,k')$, where $\mathbf{M}^h(k,k')$ is Hermitian and $\mathbf{M}^u(k,k')$ is unitary. Then in the low squeezing regime (low photon pair probability), in the instance a single pair of photons is generated localised in specific modes, the joint-spectral-amplitude (JSA) describing spectral correlations of this bi-photon state is well approximated by the corresponding component of $\mathbf M^h(k,k')$. In particular, these complex sympletic transformations have the structure,
\beq
\begin{split}
\label{eqs:polarSolution}
\mathbf M^h(k,k')&=\arFour{\alpha^h(k,k')}{\beta^h(k,k')}{\beta^{h*}(k,k')}{\alpha^{h*}(k,k')}\\
&=\exp \{i \mathcal{I}^- \mathcal{H}^{NL} \} \, \approx \mathds{1} + i \mathcal{I}^- \mathcal{H}^{NL}
\end{split}
\eeq
where the nonlinear Hamiltonian-like operator $\mathcal{H}^{NL}$, is block off-diagonal since $\mathbf M^h(k,k')$ is Hermitian and the final approximation holds for low squeezing power. In this case, the elements of $\beta^h(k,k')$ approximate $\mathcal{H}^{NL}$ which describes the generation of the bi-photon states, and we may select terms $\beta^h_{ij}(k,k')$ to associate to the JSA of bi-photons generated across modes $i$ and $j$. Note that having performed the polar decomposition,  by construction $\beta^h_{ij}(k,k')=\beta^h_{ji}(k',k)$.

We will make extensive use of the Schmidt decomposition which can be achieved by integro-singular value decomposition of the relevant functions $\beta^h_{ij}(k,k')$\com{~\cite{Law2000}}. The singular values, $\lambda_i$ (equivalently Schmidt coefficients) allow the heralded photon purity to be calculated by $P=(\sum_i \lambda_i)^4/(\sum_i \lambda^2_i)^2$. To numerically solve Eqs.~\ref{eqs:fullSolution} we discretise the wave-number $k$, sample $201$ points on the interval $k\in(-2515.01, 2515.01)\com{\text{m}^{-1}}$ (0.6 times the free-spectral range) which well resolves the features over the parameter regimes considered. Finally, we can also interrogate the temporal correlations of a bi-photon by performing the appropriate Fourier transform (FT) of the JSA to obtain the Joint-temporal amplitude JTA. We pad this Fourier transform to obtain a temporal resolution 3 times greater than the raw FT.  

\begin{figure}[ht]
\centering
\vspace*{-0.1in}
\hspace*{-0.04\columnwidth}
\includegraphics[width=1.05\columnwidth]{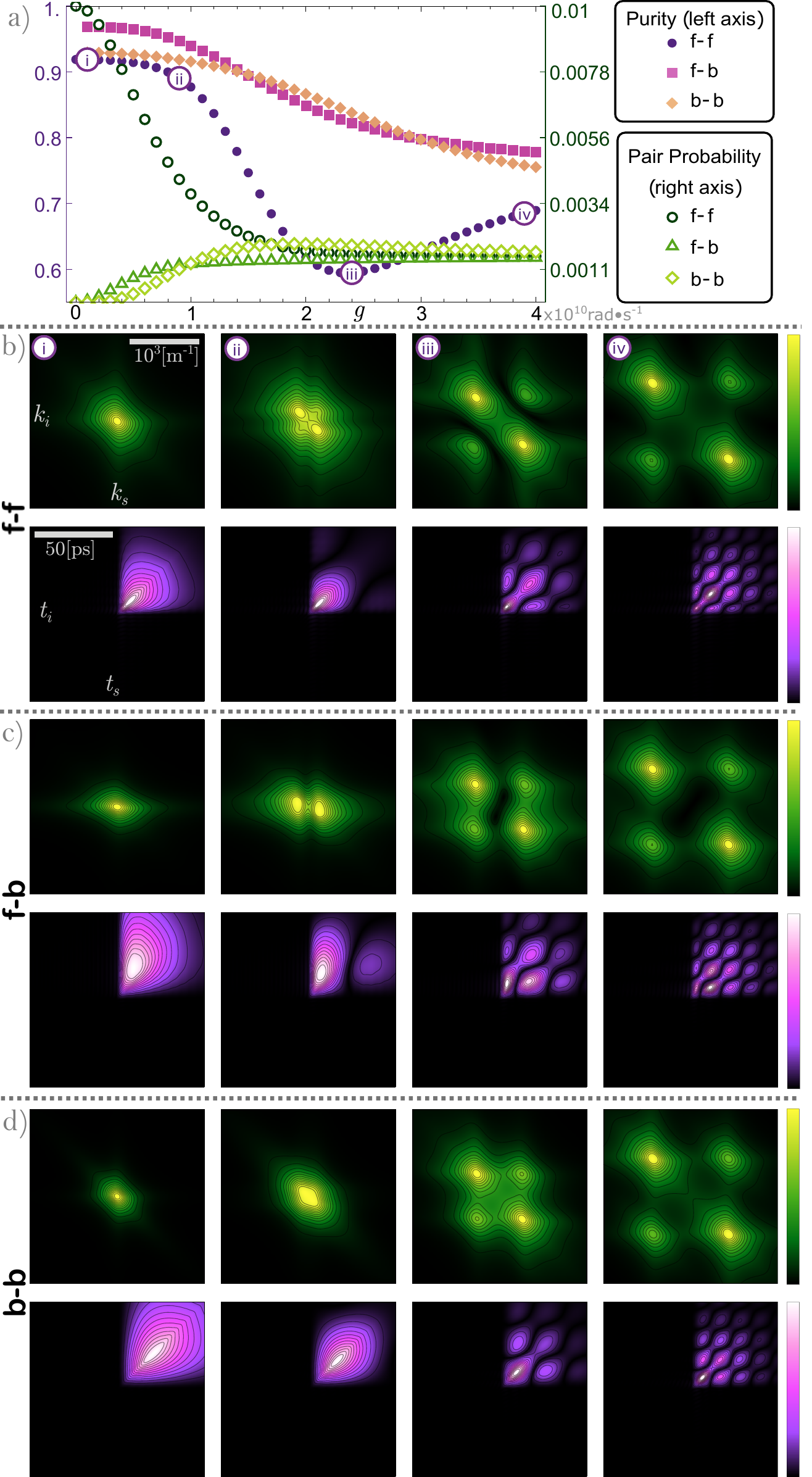}
\caption{\ref{ssec:ASplit} \textit{All Fields Splitting --- } a) Heralded photon purity and generation probability for forward-forward (\mbox{\textbf{f-f}}), forward-backward (\mbox{\textbf{f-b}}), and backward-backward (\mbox{\textbf{b-b}}) bi-photons, with all fields equally split, $g_i=g_s=g_p$.  Parts b), c), and d) contoured density plots of the JSA (green/upper) and the JTA (purple/lower) of bi-photon pairs in the \mbox{\textbf{f-f}}, \mbox{\textbf{f-b}} and \mbox{\textbf{b-b}} output modes respectively. Columns correspond to the points,\emph{i),ii),iii),iv)}, marked on a).}
\label{fig:allSplitPuritys}
\end{figure}

We focus on how the purity and the heralding efficiency of a bi-photon postselected in various output bus channels is affected by backscattering. 
We focus on three cases of interest. For each case, the Figures~\ref{fig:pumpSplitPuritys},~\ref{fig:idlerSplitPuritys} and ~\ref{fig:allSplitPuritys}, exhibit:
Part a) Heralded photon purity (left axis) and generation probability (right axis), Parts b), c), and if exists d), contoured density plots of the JSA (green/upper) and the JTA (purple/lower) of bi-photon pairs in specified output buses with columns corresponding to the points marked on a) with \emph{i),ii),iii),iv)}. The JSA and JTA axes span $k\in(-1257.51, 1257.51)\text{m}^{-1}$ and $t\in (-55.64,55.64)\text{ps}$ respectively. The pair probability (sub-figure a) indicates the square integral over these densities, so the displayed values are each normalised for clarity with the relevant colour function legend positioned to right.

\subsubsection{Pump Splitting}
\label{ssec:PSplit}
When only the pump field exhibits splitting, the forward and backward propagating modes of the signal and idler are independent and depend only on the corresponding (forward and back) propagating pump fields.  Furthermore, the bus and loss modes are equivalent, in that, all squeezing between loss and bus modes exhibits the same spectral properties, a feature that survives even outside of the critically coupling regime up to a constant factor. There are then, just two relevant two-mode squeezing terms to consider in $\mathbf{M}^h(k,k')$, those corresponding to photon pairs that are produced either both propagating forward (\textbf{\mbox{\textbf{f-f}}}) or backward (\mbox{\textbf{b-b}}). 

\begin{figure}[h]
\centering
\vspace*{-0.1in}
\hspace*{-0.04\columnwidth}
\includegraphics[width=1.05\columnwidth]{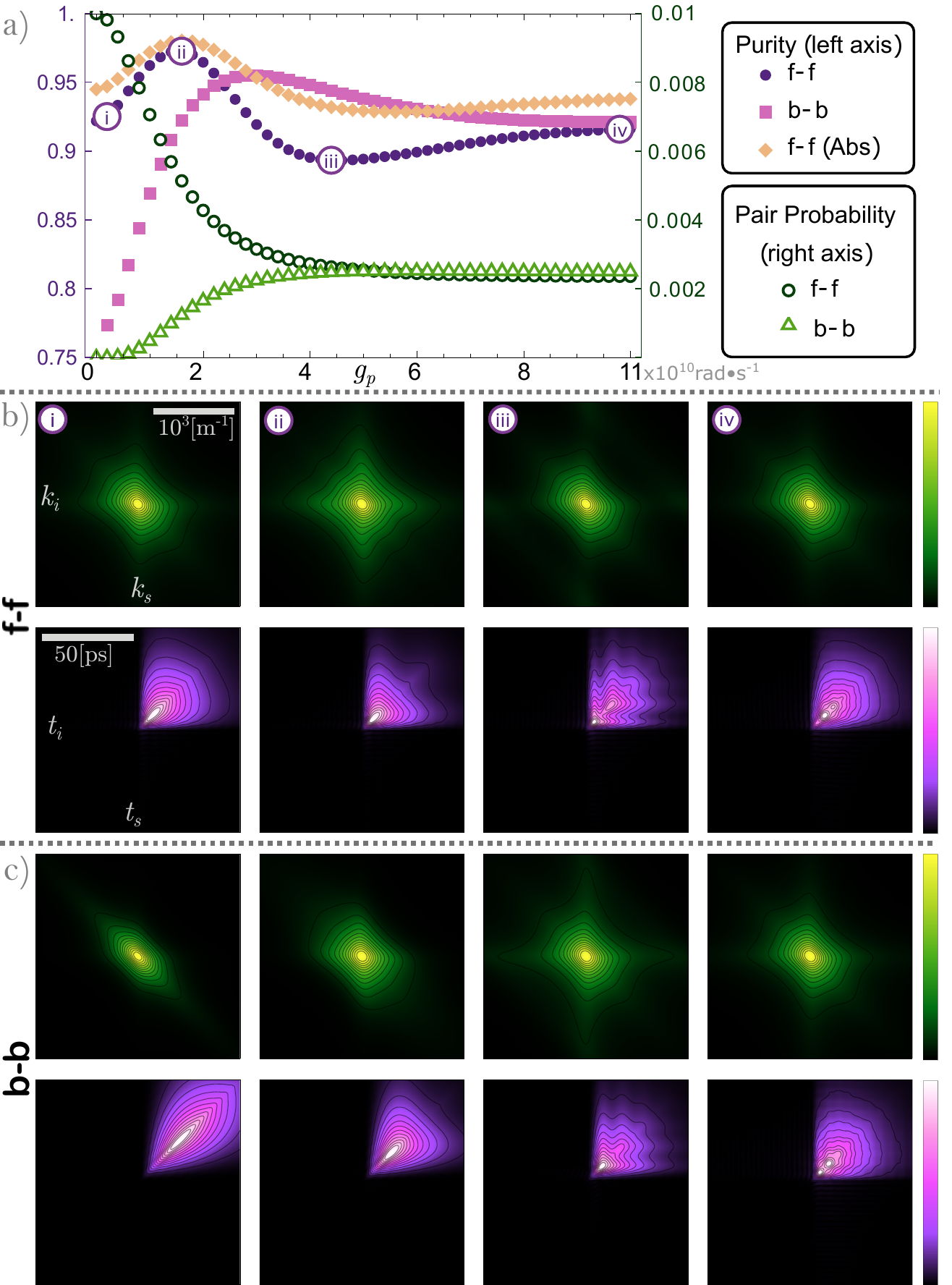}
\caption{ \ref{ssec:PSplit} \emph{Pump Splitting} --- a) Purity and pair probability with pump mode splitting $g_p$, for forward-forward (\mbox{\textbf{f-f}}) and backward-backward (\mbox{\textbf{b-b}}) generated photon pairs. Also the purity implied by the absolute value of the \mbox{\textbf{f-f}} JSA.  Parts b), c), contoured density plots of the JSA (green/upper) and the JTA (purple/lower) of bi-photon pairs in the \mbox{\textbf{f-f}} and \mbox{\textbf{b-b}} output modes respectively. Columns correspond to the points,\emph{i),ii),iii),iv)}, marked on a).}
\label{fig:pumpSplitPuritys}
\end{figure}

In Fig.\ref{fig:pumpSplitPuritys}, \textbf{a)} and \textbf{b)}, we observe an increase in purity for the \mbox{\textbf{f-f}} pairs as the backscattering parameter for the pump, $g_p$, is increased, attaining a maximum of $\approx0.97$, followed by a decrease to $\approx0.89$ and a gradual return to the raw purity. The increase at small $g_p$ can be attributed to an effective broadening of the pump spectrum, which loosens the energy matching constraint which in standard microrings leads to correlations in the signal-idler JSA, and thus heralded photon impurity. A gentle decrease in efficiency can be seen through the drop in pair rate as the peak pump power in the ring decreases. As splitting gets very large the convolution of the pump's spectrum becomes dominated by the central of the three peaks (see Fig.~\ref{fig:pumpSplittingLinear}) and the system approaches the non-degenerately pumped case, in which two independent Lorentzian cavity modes drive the process, which is analogous to the non-split case (providing a broad enough injected pump) all but for a factor of $1/4$ in the generation rate arising from a combinatorial factor of the process. In the intermediate region, \emph{iii)}, whilst the absolute value of the JSA closely resembles the non-split case (corroborated by the purity of the absolute value of the JSA), the correlations in the phase reduce the purity below the non-split case, which can be readily observed in the JTAs, whereby the secondary peak along $t_s=t_i$ indicates a suppression and subsequent revival of the forward propagating pump power as the forward and backward cavity modes beat together. This observation helps clarify additional phase-dependency of the system, and demonstrates that increased pump broadening by splitting is not sufficient alone to obtain photon purities beyond 0.97. 

Despite the apparent symmetry in the parameters of the set-up, the \mbox{\textbf{b-b}} propagating pairs, depicted in part \textbf{c)}, behave very differently due to pump being injected solely in the forward propagating bus mode, and they are consequently nonexistent at $g_p=0$ (due to the absence of the back propagating field). For small $g_p$, column \emph{i)}, the \mbox{\textbf{b-b}} pairs exhibit strong correlations with minimum purity $0.77$ owing to the back-propagating pump having a narrower spectrum, in part due to the filtering it must undergo to couple through the forward propagating cavity mode. Nonetheless, broadening of the pump as $g_p$ increases still leads to purities in excess of 0.95, though these reduce to the non-split ring purities with the onset of the non-degenerately pumped regime. 

\subsubsection{Idler Splitting}
\label{ssec:ISplit}
\begin{figure}[h]
\centering
\vspace*{-0.1in}
\hspace*{-0.04\columnwidth}
\includegraphics[width=1.05\columnwidth]{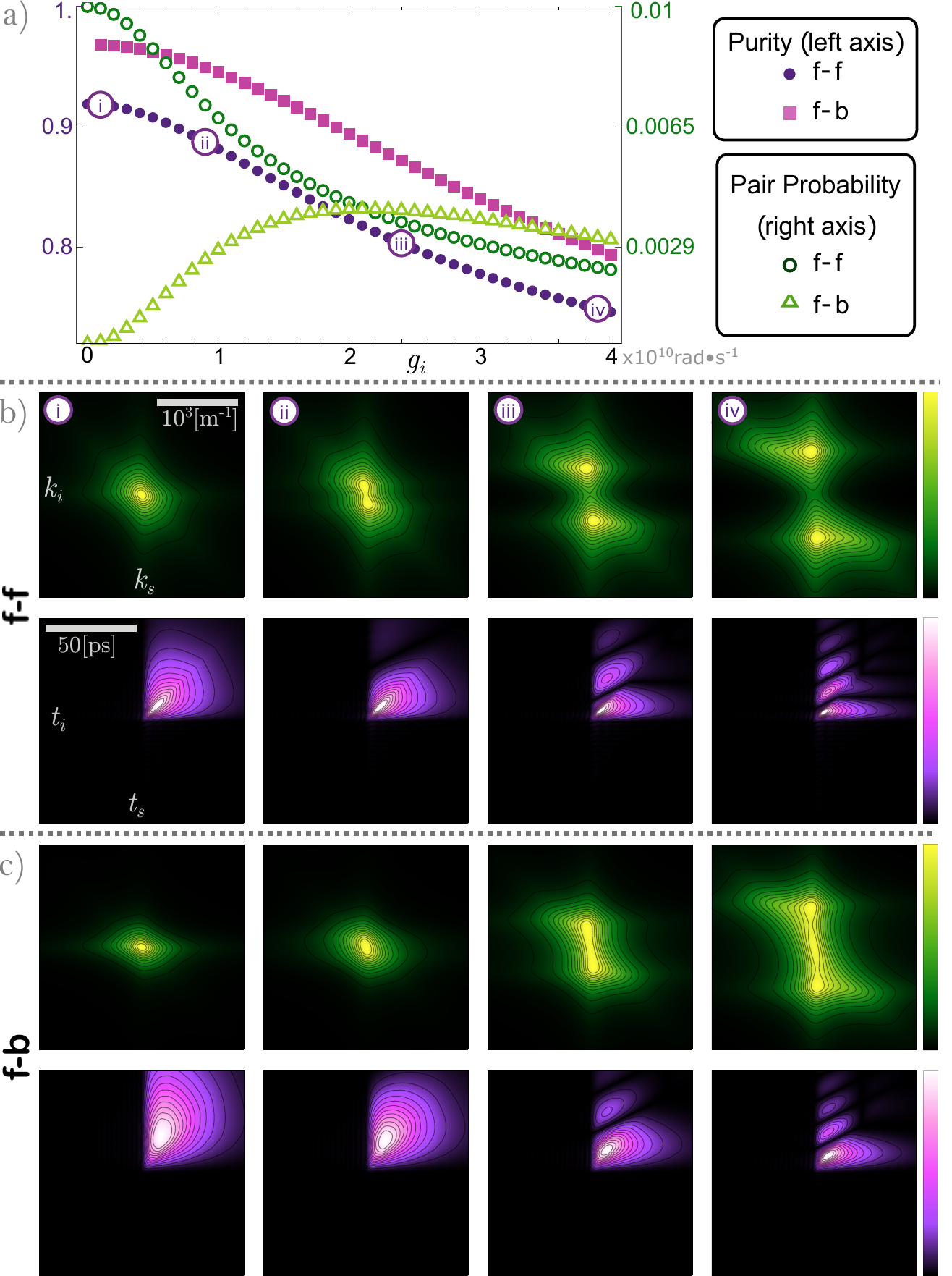}
\caption{\ref{ssec:ISplit} \emph{Idler Splitting ---} a) Heralded photon purity and generation probability for forward-forward (\mbox{\textbf{f-f}}) and forward-backward (\mbox{\textbf{f-b}}) with Idler mode splitting $g_i= \{1/10, 9/10, 12/5, 39/10 \}\times 10^{10}$\com{\radss}. Parts b), c), contoured density plots of the JSA (green/upper) and the JTA (purple/lower) of bi-photon pairs in the \mbox{\textbf{f-f}} and \mbox{\textbf{f-b}} output modes respectively. Columns correspond to the points,\emph{i),ii),iii),iv)}, marked on a). The \mbox{\textbf{f-f}} photons exhibit reduction in both purity and pair probability with increased $g_i$. The double peaked idler spectrum becomes quickly apparent in the JSA, accompanied by beating in the JTA which increases in frequency as the coupling becomes stronger. The \mbox{\textbf{f-b}} pairs however, attain purity of $0.968$ for small $g_i$ owing to the effective tighter filtering of the idler through the back-propagating mode evidenced in the narrow JSA, c)\emph{i)}. For larger splitting, since the backward out-coupled idler field enhancement doesn't vanish at $k=0$, unlike the forward propagating case (see Fig.~\ref{fig:pumpSplittingLinear}, \emph{ii)}, for an analogous case) the generated idler state exhibits a non-vanishing intensity in the spectrum between the peaks of the split resonance.}
\label{fig:idlerSplitPuritys}
\end{figure}

With only idler mode splitting, all photons are generated by the forward propagating pump field, so the signal field is non-zero only in the forward propagating direction, though the idler scattering allows for pairs to be produced with the idler exiting in the backward mode (\mbox{\textbf{f-b}}), as well as the forward mode (\mbox{\textbf{f-f}}).
In Fig.~\ref{fig:idlerSplitPuritys} we see the purity and pair probabilities for the case where only the idler mode is split by parameter $g_i$. The purity of the heralded photon suffers as a result of splitting, and these effects can be readily seen observing the changes to the \mbox{\textbf{f-f}} JSAs. In the case the forward propagating idler mode is used as herald, the heralding efficiency is unchanged by the splitting since every forward propagating herald is accompanied by a heralded signal photon (up to the 0.5 probability caused by the critically coupled phantom/loss channel). If however the signal is used as the herald, some partner photons are lost into the back-propagating modes so the heralding efficiency would be reduced by a factor given by the ratio between the \mbox{\textbf{f-f}} and \mbox{\textbf{f-b}} pair probabilities. These observed single photons (forward propagating signal photons without accompanied idler photons) can be understood as a thermal source when tracing over the back-propagating idler mode, but unlike in the case of loss applied to a two-mode squeezed source, this thermal contribution is characterised by an altogether different spectrum. This spectrum, given by the marginal of the \mbox{\textbf{f-b}} JSA \com{(where the idler marginal is obtained by $\int d  k_s F(k_s,k_i)F^*(k_s,k_i)$)}, could in principle, be filtered or mode sorted to some degree, from the \mbox{\textbf{f-f}} JSA of the bi-photon term.

In the case of the \mbox{\textbf{f-b}} bi-photons the pair probability quickly approaches appreciable levels for even relatively slight splitting. The purity of such \mbox{\textbf{f-b}} bi-photons begins at $0.968$ at very low rates, owing to the narrower filtered spectrum of the idler photons coupling out backwards, though reduces as the splitting increases and the states become brighter. At large splitting the \mbox{\textbf{f-b}} JSA differs significantly from the \mbox{\textbf{f-f}} case, in that it's non-vanishing at $k_s=k_i=0$, which can be seen to result from the non-vanishing field enhancement of the idler at $k_i=0$, analogous to the back propagating pump mode in Fig.~\ref{fig:pumpSplittingLinear}, \emph{iii)} and \emph{iv)}.

\subsubsection{All Fields Splitting}
\label{ssec:ASplit}
When all fields are equally split ($g_p=g_i=g_s=g$) we can observe photon pairs in all configurations of forward and back propagating and we consider the cases, \mbox{\textbf{f-f}}, \mbox{\textbf{f-b}}, and \mbox{\textbf{b-b}}. The purity of the \mbox{\textbf{f-f}} bi-photons decreases less quickly than in the pure idler splitting case, owing to the pump, and thus energy-matching constraint, broadening. For large $g$ the JSA tends to a four peaked function, whereby the loose energy matching constraints, owing to the broad pump, allow photons to be generated in all four combinations of energies of the bi-peaked signal and idler resonances and the JTA exhibits symmetric beating. In the \mbox{\textbf{f-b}} case, the asymmetry is clear in the JSA and JTA for small $g$ though for larger $g$ the JSA appears not dissimilar to the \mbox{\textbf{f-f}} case, though the JTA still features this asymmetry through a tendency towards larger delays in the idler. The \mbox{\textbf{b-b}} case retains the symmetry of the \mbox{\textbf{f-f}} case, though the JTAs show the tendency towards larger delays due to the slower excitation of the back-propagating pump mode due to coupling via the forward mode.  

\begin{figure}[h]
\centering
\vspace*{-0.1in}
\hspace*{-0.04\columnwidth}
\includegraphics[width=1.05\columnwidth]{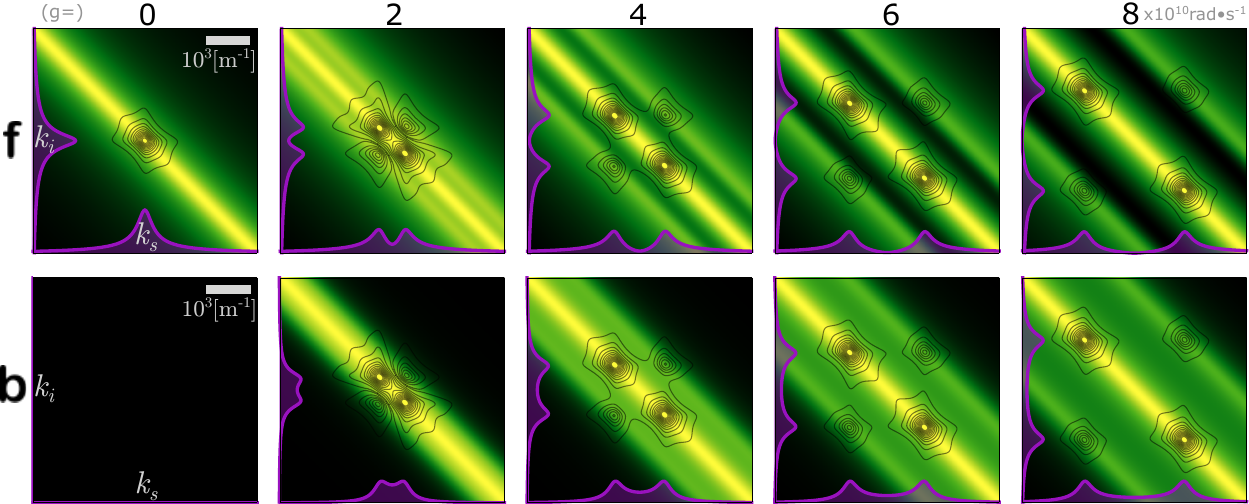}
\caption{The contour plots display the output JSAs in the case of equal splitting for the range of $g=\{0,2,4,6,8\}\times 10^{10}$\com{\radss}. The density plots (green) display the relevant (\mbox{\textbf{f-f}}) term from ${\bar \Gamma}^{\text{Sq}}(k,k')$, whilst the axes plots (purple) show the linear transmission from the given ring mode to the output bus, $L^{a \rightarrow s^+}$ (Appendix.~\ref{app:pertubative}). Each is plotted over the full simulated range $k\in(-2515.01, 2515.01)\com{m^{-1}}$. }
\label{fig:allSplitOverlays}
\end{figure}

To better understand the origins of these features we turn to the perturbative solution given in Appendix~\ref{app:pertubative}, whereby the contributions to the JSA can be understood as arising from the nonlinear part of ${\bar \Gamma}^{\text{Sq}}(k,k')$ being resonantly filtered to the output ports. In this case, the JSA is given by the coherent sum of the contributions from the forward and backward propagating modes in the ring. In Fig.~\ref{fig:allSplitOverlays} we display the \mbox{\textbf{f-f}} JSA contour plot with the ${\bar \Gamma}^{\text{Sq}}(k,k')$ density plot overlaid and the signal and idler resonant filtering functions on the axes, for a (wider) range of all equal splitting parameter, $g$. The sum of the forward (top) and backward (bottom) terms constitute a perturbative explanation of the observed JSAs. 

\section{Stochastic Methods}
\label{sec:stochastic}
\begin{figure}[h]
\centering
\vspace*{-0.1in}
\hspace*{-0.04\columnwidth}
\includegraphics[width=1.05\columnwidth]{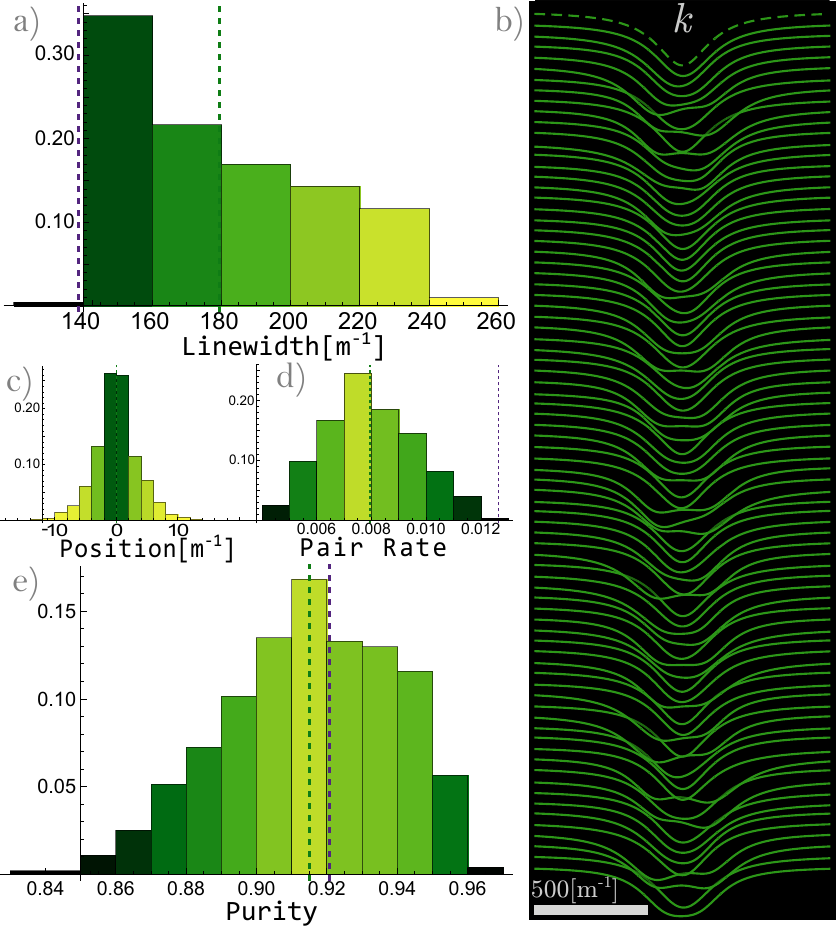}
\caption{a), c), Histograms of the linewidth and position of the 3000 resonances from the stochastic sample (Section~\ref{sec:stochastic}). The mean linewidth (green) of the sample is $\delta k = 179\com{m^{-1}}$ compared to just $139$ in the case of no backscattering (purple). b) \com{The transmission line shape with no backscattering (dashed) and 80 random resonances from the sample (solid).} Whilst some show splitting, most appear to display just slight broadening, characteristic of often observed transmission spectra. d) and e) show histograms of the pair rate and purity of the 1000 photon pair sources (comprised of 3 such resonances). The pair rate is always reduced from the case with no backscatter $0.0124$ (purple), with the mean $0.00795$ (green). The photon purity suffers only a minor averge reduction from $0.921$ (purple) to $0.915$ (green) with a significant proportion of sources actually displaying an increased purity, with the maximum achieving $0.964$. As discussed in the main text however, the ensemble purity is significantly reduced.}
\label{fig:StochasticPlots}
\end{figure}

Whilst the trends for particular forms of backscattering are insightful and could inform the design of \com{systems of coupled cavity modes}, in practice, fabrication defects are out of the experimenters control and behave randomly from sample to sample, and resonance to resonance\footnote{The finite length of the ring implies a scattering correlation length of $\lambda^2/2n_g L_r \approx 1.31125\text{nm}$~\cite{Morichetti2010b} making each resonance independent though scattering uniform over a given resonance.}. The often observed slight asymmetric splitting example of Fig.~\ref{fig:pumpSplittingLinear} a), invites an analysis of the statistics of resonators with qualitatively similar transmission properties. We assess the properties of an ensemble of 1000 devices having each resonance's (pump, signal, and idler) \com{parameters sampled from uniform distributions over the intervals} $|g|\in(0,10^{10})$\com{\radss}, $|\delta_f|\in(0,0.2)$, $|\delta_b|\in(0,0.2)$ and $|c|\in(0,0.2)$\com{\radsm}, with the phase of each random. In Fig.~\ref{fig:StochasticPlots} a) a histogram of the observed linewidths is presented and part b) shows the first 80 transmission line shapes \com{(solid) of such cavity modes, along with the ideal case (dashed), to provide a qualitative understanding of the resultant line broadening}. Whilst these parameters are not significant enough to often show splitting of the resonance, some line broadening is clearly present in most cases as well as slight wandering of the mean position of the resonances as shown in c). The mean purity of the \mbox{\textbf{f-f}} bi-photons observed from this ensemble was $0.915$, only slightly reduced over the ideal (non-split) case at $0.921$, as presented in the histogram Fig~\ref{fig:StochasticPlots} e). The mean pair probability however drops to $0.00795$ from the non-split case $0.01265$(Fig~\ref{fig:StochasticPlots} d) ). 

Despite this relatively minor reduction in the source purity and probability, the quality of such heralded photons is often only useful when many such sources show high quality interference between one another, which invites consideration of the ensemble purity of these sources, ie. the purity of the mixed state comprised of an equally weighted mixture of the heralded single photons from each. We find ensemble purity of $0.8867$ indicating that the visibility of interference from separate sources suffers more significantly. 

\begin{figure}[h!]
\centering
\vspace*{-0.1in}
\hspace*{-0.04\columnwidth}
\includegraphics[width=1.05\columnwidth]{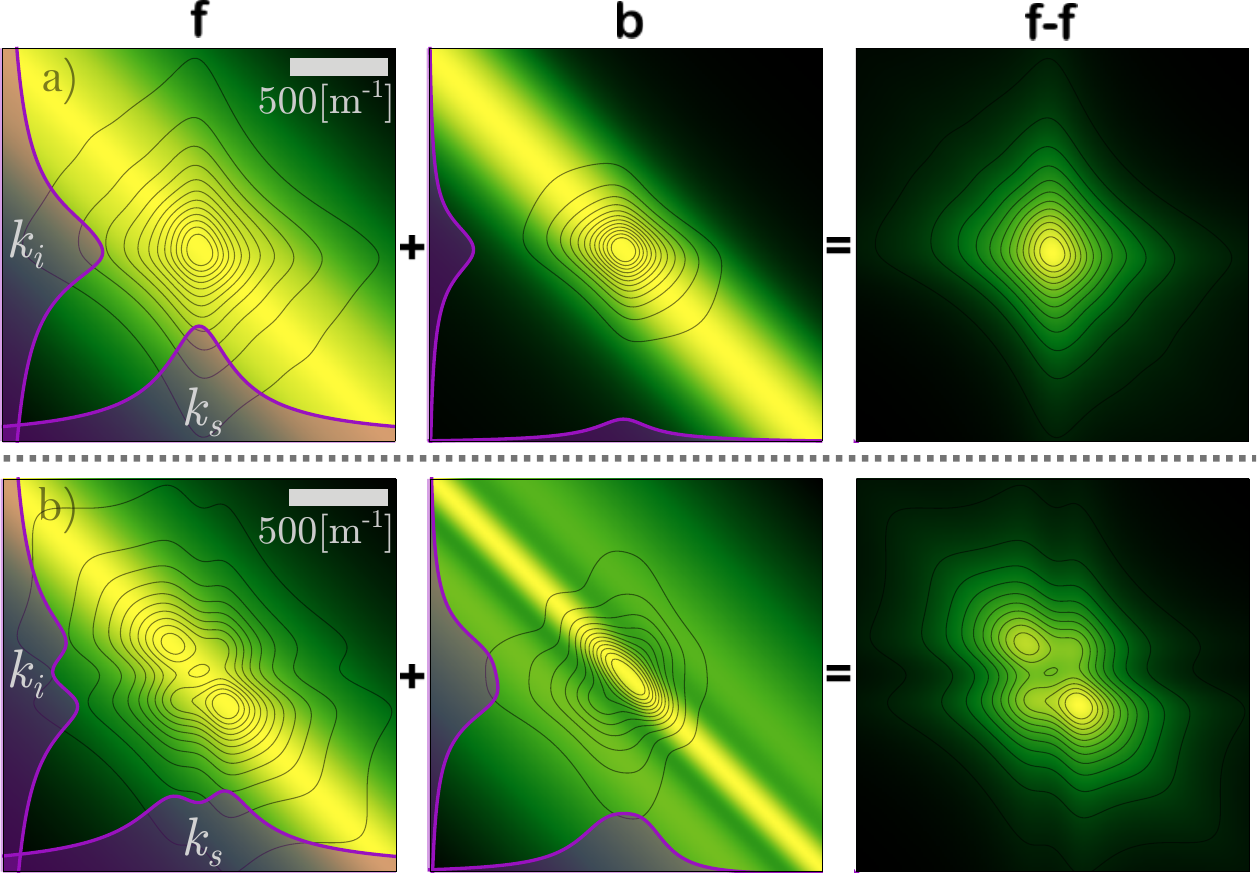}
\caption{Parts a) and b), referring to the best and worse case purities respectively, show the two contributions, forward generated (left), and backward generated (central), to pairs output in the forward-forward configuration (right). The density plots (green) display the relevant pump fields convolution from ${\bar \Gamma}^{\text{Sq}}(k,k')$, whilst the axes plots (purple) show the linear transmission from the given ring mode to the output bus, $L^{a \rightarrow s^+}$, resulting in the contour plot as per Appendix.~\ref{app:pertubative}. Each is plotted over $k\in \{-1006.01, 1006.01\}\com{m^{-1}}$.}
\label{fig:BestWorstPerturbativeJSA}
\end{figure}

A discussion of the best and worst cases is in order. The perturbative terms contributing to the best and worst case JSAs are depicted in Figure.~\ref{fig:BestWorstPerturbativeJSA}, a) and b) respectively. The best case, achieves a purity of $0.964$, slightly short of the optimum purity, $0.968$, achieved by the pure pump splitting case (Section~\ref{ssec:PSplit}). Figure.~\ref{fig:BestWorstPerturbativeJSA}, a) demonstrates that narrow signal and idler resonances (small backscattering parameters), along with a broad pump resonance lead to higher purities as with the pump split case. Whilst of course this is circumstantial, it provides evidence that purities in excess of this may not be possible for systems of this form. Though it is worth noting that the form of the coupling terms we have imposed to model backscatter, need not hold in more general \com{systems of coupled cavity modes} and \com{it would certainly be of interest to consider different coupling models\cite{Vernon2017} and explore the design of optimal sources under these kinds of defects}. The worst case, on the other hand, has a purity of just $0.837$, despite the relatively minor splitting parameters. In Figure.~\ref{fig:BestWorstPerturbativeJSA}, b) we see relatively broad, and clearly split, signal and idler resonances along with a narrow pump resonance. We remark that these perturbative methods obtain fidelities $0.999917$ and $0.999531$ to the full solutions, for the best and worst cases respectively. This is to be expected since the full $\beta^h$ takes the form of a hyperbolic sine in $\Gamma^{(Sq)}$, and the truncated series omits terms beyond $\mathcal{O}(\Gamma^{(Sq)3})$, whilst the pair rate satisfies $|\Gamma^{(Sq)}_{ij}|_F\approx 0.01$. 

\section{Challenges for Stimulated Emission Tomography}
\label{sec:SET}
Stimulated emission tomography (SET), a common method for performing tomography of such systems, consists of inserting a CW seed laser to drive the signal (idler) mode and recording the spectrum of the resultant conjugate beam in the idler (signal)~\cite{Liscidini2013,Borghi2020,Triginer2020c}. Having pursued a Gaussian treatment, this behaviour is directly described by the transformation Eqs.~\ref{eqs:fullSolution}~\cite{Serafini2017a} providing the undepleted pump approximation still holds, i.e. the pump fields are not significantly perturbed by displacements of the same order of the conjugate beam displacements. 

The standard prescription then consists of the following reasoning: measure elements of the full transformation, Eqs.~\ref{eqs:fullSolution}, and use these to infer elements of the Hermitian part of the polar decomposition of the transformation, Eqs.~\ref{eqs:polarSolution}, which describes the spontaneous (no seed laser) behaviour of the source. In the instance the entire transformation (Eqs.~\ref{eqs:fullSolution}) is measured, the polar decomposition is straightforward. Similarly, if sufficient structure is assumed in Eqs.~\ref{eqs:fullSolution}, (such as, for instance, the symmetry between the bus and loss channels in the case of no backscatter) it may be possible to infer properties of the entire transformation, and subsequently its polar decomposition, from incomplete data, ie. in the absence of any characterisation involving seeding or measuring the loss channels. Assumptions of this kind however, are not available in the case of backscattering, ie. SET measurements involving only seeding and measurement of the bus channels, do not uniquely determine the spontaneous behaviour of the source. This becomes apparent when considering that the spontaneous behaviour of the source depends on squeezing of the vacuua that enter from various ports of the device, without sufficient knowledge or assumptions, the squeezing applied to the vacuum entering inaccessible inputs to the device (whilst exiting in the forward bus mode) cannot be determined.

\begin{figure}[h!]
\centering
\vspace*{-0in}
\hspace*{-0.0\columnwidth}
\includegraphics[width=1.0\columnwidth]{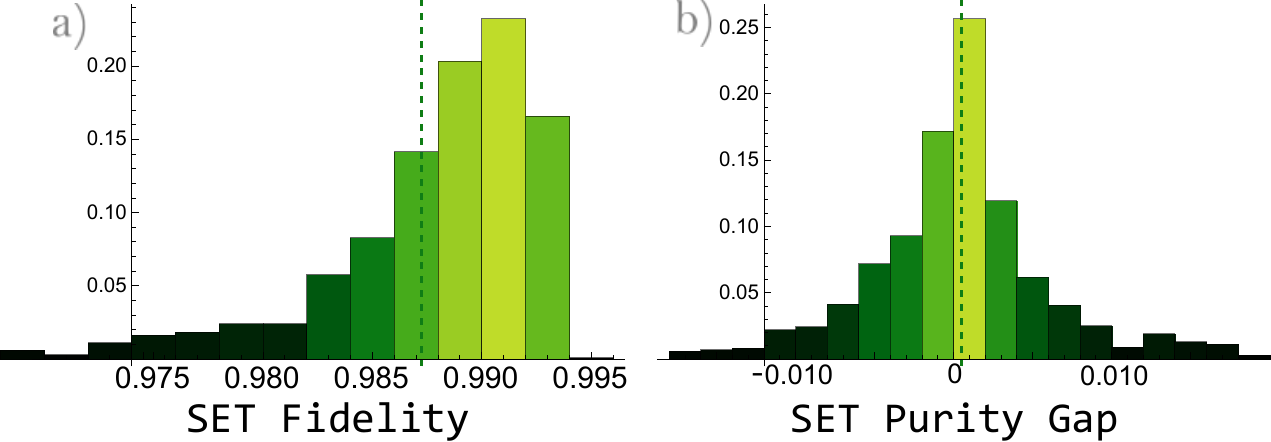}
\caption{a) Histogram of the fidelity between the spontaneous \mbox{\textbf{f-f}} bi-photon state and the inferred bi-photon state from the standard stimulated emission tomography procedure, over the 1000 samples of the stochastic ensemble of Section~\ref{sec:stochastic}. b) Histogram of the difference between the inferred purity and true purity of the stochastic ensemble.}
\label{fig:StochasticSET}
\end{figure}

Were we to assume no backscattering was present, and estimate the spontaneous JSA using the conventional method (outlined in Appendix~\ref{app:SETPolar}), our tomography miss-estimates the outgoing state generated in the spontaneous regime, and a histogram of the fidelity of this estimation procedure (over the ensemble, Section~\ref{sec:stochastic}) is provided in Figure~\ref{fig:StochasticSET} a). We find a mean fidelity of $0.9872$ is obtained, which puts strict limits the applicability of such tomography schemes when moving towards source purities near or exceeding this. 
Finally, we consider whether such an approach to tomography biases our inferred purity towards over- or under-estimating the quality, and present the purity gap (difference in the in SET inferred purity and the actual purity) in Figure.~\ref{fig:StochasticSET}d). The mean purity gap is just $0.00049$, so whilst we suspect any bias is insubstantial over this particular ensemble, it is by no means apparent how other source defects, or particular distributions over them, may affect tendencies toward infidelity of this estimation process. 

In light of these challenges, it is clear that to achieve high fidelity assumption free tomography of spontaneous Gaussian processes, including single photon sources, one requires additional data or assumptions, including synthesis of linear properties of the system or seeding and detection of the additional modes at play.

\section{Discussion}
The resultant transformations demonstrate an important feature of squeezing in the presence of multiple spectral and spatial modes - the states of light exiting the device in some modes of interest (the forward propagating signal and idler modes in the bus for instance) are no longer fully characterised by the single JSA~\cite{Helt2015}, \com{nor do a single pair of joint-spectra from stimulated emission uniquely define the transformation.} Instead, squeezing between different spatial modes has consequences for the modes of interest. Furthermore, whilst the heralded bi-photon picture of events applies readily in the low squeezing limit, or where all modes (including loss modes) could be monitored with perfect detection, in general multi-pair events muddy any attempt to isolate a single bi-photon and require more involved analysis. Finally, an unfortunate disparity has occurred in this process, whereby the Schmidt decomposition of the bi-photons we have considered, no longer corresponds to a Bloch-Messiah decomposition of the underlying Gaussian process~\cite{Ma1990,Wasilewski2006,Braunstein2005}. This lack-of correspondence inhibits to a large extent, how multi-photon behaviour can be efficiently modelled in such systems, and whilst some progress in this direction has been made~\cite{Thomas2020,McCutcheon2018}, we leave the complete analysis to future work, where we expect the nature of multiphoton terms takes on a rich structure.  

\section{Conclusion}
\com{We have derived a comprehensive Hamiltonian based input-output model for Gaussian optical processes in the presence of multiple cavity modes for general coupling.
This framework can be readily applied to arbitrary systems of coupled cavity modes, and allows the exploration of a wide variety of future quantum light sources. 
The application of such a framework involves just the input of the systems parameters and the inversion of linear equations. 

This was directly applied to the case of a single microring exhibiting backscatter, backcoupling, and backreflection, with relevant  Silicon-on-insulator waveguide properties. This model, more entire than previous treatments, reproduces observed asymmetric linear phenomena, and the joint-spectra of the photon pair generated by four-wave mixing in such rings was evaluated for various simple backscattering regimes. The limitations of closed form solutions to such systems have been expressed, and a perturbative solution has been presented shedding light on the behavior of such systems for small nonlinearities. 

Additionally, the properties of sources sampled from a representative distribution of parameters was considered, an observed slight decrease in single source purity, whilst a significant drop in ensemble purity was observed. The consequences for stimulated emission tomographic methods were highlighted. This all aids in our understanding of the consequences of backscattering in photon sources, helps inform the development of applicable methods to tomograph these systems, and ultimately how to mitigate the problem of fabrication defects in cavity based quantum light sources.  

One further example of the application of this frameworks was considered involving a photonic molecule, demonstrating the straightforward nature of its generalization. Finally, the structure of the obtained transformations invites the consideration of more in-depth analysis of the effects of multi-photon noise in quantum photonic systems via a Gaussian process framework.}

\section*{Acknowledgements}
The author would like to thank Prof. J.G.Rarity, Drs G.F.Sinclair, J.Silverstone,  M. Borghi, D.P.S.McCutcheon, O.Thomas for fruitful discussion. This work was supported by the UK's Engineering and Physical Sciences Research Council (EP/P510269/1 and EP/L024020/1). The author has no competing interests.

\section*{Data Availability}
The data that support the findings of this study are available from the corresponding author upon reasonable request.

\newpage
\onecolumngrid
\renewcommand{\theequation}{\thesection.\arabic{equation}}
\setcounter{equation}{0}
\begin{appendices}
\section{Channel Equations of Motion}
\label{app:inputOutput}
The Heisenberg equations of motion for the channel fields are,
\beqs
\begin{aligned}
(\frac{\partial }{ \partial t}+ v_j \frac{\partial}{\partial z} +i \Omega_j ) \hat s_j(z,t) =  -i \gamma^*_{nj} \hat a_n(t)\delta(z) - i C_{jl} \hat s_l(0,t) \, ,
\end{aligned}
\eeqs
or in vectorised notation,
\beqs
\begin{aligned}
(\frac{\partial }{ \partial t}+ V \frac{\partial}{\partial z} +i \Omega ) \mathbf s(z,t) =  -i \gamma^\dagger \mathbf a(t)\delta(z) - i C \mathbf s(0,t) \, ,
\end{aligned}
\eeqs
which can be solved by a coordinate transform $\xi_j=x-v_j t$
admitting the general solution \cite{Vernon2015},
\beqs
\begin{aligned}
\hat s_j(z,t) = \mathcal{S}_j(z,t) -\tfrac{i}{v_j} \biggl( \gamma^*_{nj} \hat a_n(t-z/v_j) + C_{jl} \hat s_l(0,t-z/v_j) \biggr) [\Theta(z) - \Theta(z-v_j/t)]\exp\{-i \Omega_j z/v_j\} \, ,
\end{aligned}
\eeqs
with $\mathcal{S}_j(z,t)$ any solution to the homogeneous problem. The one-sided limits read,
\beqs
\begin{aligned}
\hat s^-_j(0,t) &:= \lim_{z\rightarrow 0^-} \hat s_j(z,t)= \mathcal{S}_j(0,t)\\
\hat s^+_j(0,t) &:= \lim_{z\rightarrow 0^+} \hat s_j(z,t) = \mathcal{S}_j(0,t)-\tfrac{i}{v_j} \biggl( \gamma^*_{nj} \hat a_n(t) + C_{jl} \hat s_l(0,t) \biggr)\\
\end{aligned}
\eeqs
At the coupling region we will impose that the discontinuity obeys,
\beqs
\begin{aligned}
\hat s_j(0,t) &= \half \bigl ( \hat s^-_j(0,t)+\hat s^+_j(0,t)  \bigr)\\
&=\hat s^-_j(0,t)-\tfrac{i}{2 v_j} \biggl( \gamma^*_{nj} \hat a_n(t) + C_{jl} \hat s_l(0,t) \biggr)\\
(\mathds{1}_{jl} + \frac{i}{2v_j} C_{jl})\hat s_l (0,t) &= \hat s^-_j(0,t)-\tfrac{i}{2 v_j} \gamma^*_{nj} \hat a_n(t) \\
\Rightarrow \hat s_l (0,t) &= \tilde C^{-1}_{jl} \biggl(\hat s^-_l (0,t) - \frac{i}{2v_l} \gamma^*_{nl} \hat a_n (t) \biggr) \\
\text{with} \quad  \tilde C_{jl} &:=(\mathds{1}_{jl} + \frac{i}{2v_j} C_{jl})
\end{aligned}
\eeqs
and derive the input-output relations,
\beqs
\begin{aligned}
\hat s^+_j(0,t)&=\bigl(\mathds{1}+\frac{i}{2} V^{-1}C\bigr)^{-1}_{jo}\biggl(\bigl(\mathds{1}-\frac{i}{2}V^{-1}C\bigr)_{ol}\hat s^-_l(0,t) - \tfrac{i}{v_j} \gamma^*_{nj} \hat a_n(t)\biggr) \\
\end{aligned}
\eeqs

\section{Perturbative Solutions}
\label{app:pertubative}
Splitting $\mathbf{\bar \Gamma}(k,k')$ into linear and nonlinear contributions,
\beqs
\begin{aligned}
\mathbf{\bar \Gamma}(k,k')&:= \mathbf{\bar \Gamma}^{(L)}(k,k')+\mathbf{\bar \Gamma}^{(NL)}(k,k')
\end{aligned}
\eeqs
where $\bm{\bar \Gamma}^{(NL)}(k,k')$ is small and can be treated as a perturbation, we can expand the inverse in the solution to Eqs.~\ref{eqs:FourierNonLinear} using a Neumann series to obtain the perturbative solution,
\beqs
\begin{aligned}
\label{eqs:FourierNonLinearPerturbativeSolution}
\int  d k'   \biggl(-i \bm{\mathcal{V}}^- k' \delta(k-k')   + \bm{\bar \Gamma}^{(L)}(k,k')+\bm{\bar \Gamma}^{(NL)}(k,k') \biggr)  \tilde {\mathbf A} (k')&=-i  \int   d k'  \bm{\bar \gamma}(k,k')      {\mathbf S}^- (k') \\
  \biggl(-i \bm{\mathcal{V}}^- k' \delta   + \bm{\bar \Gamma}^{(L)}+\bm{\bar \Gamma}^{(NL)} \biggr)  \tilde {\mathbf A} &=-i  \bm{\bar \gamma}    {\mathbf S}^- \\
  \end{aligned}
  \eeqs
  \beqs
  \begin{aligned}
  \rightarrow \,  \tilde {\mathbf A}
  &= \biggl(-i \bm{\mathcal{V}}^- k' \delta   + \bm{\bar \Gamma}^{(L)}\biggr)^{-1} \sum_{j=0}^\infty 
  \biggl(-\bm{\bar \Gamma}^{(NL)} \bigl(-i \bm{\mathcal{V}}^- k' \delta   + \bm{\bar \Gamma}^{(L)}\bigr)^{-1} \biggr)^j
  (-i\bm{\bar \gamma})    {\mathbf S}^- \\
   &= \biggl(-i \bm{\mathcal{V}}^- k' \delta   + \bm{\bar \Gamma}^{(L)}\biggr)^{-1} 
  \biggl(\mathds{1}+\bm{\bar \Gamma}^{(NL)} \bigl(-i \bm{\mathcal{V}}^- k' \delta   + \bm{\bar \Gamma}^{(L)}\bigr)^{-1} +\mathcal{O}(\bm{\bar \Gamma}^{(NL)2})\biggr)
  (-i\bm{\bar \gamma})    {\mathbf S}^- 
\end{aligned}
\eeqs
where we drop the explicit $k$ dependencies and integration for brevity. \comtwo{In the most trivial of circumstances}, this first order solution can be seen to be equivalent to solving the linear equations of motion for one of the fields (signal or idler) and using this as a source term for the nonlinear contribution in the conjugate field (idler or signal respectively). The resultant input-output transformation becomes just,
\beq
\begin{aligned}
\label{eqs:perturbativeInputOutput}
\tilde{\mathbf{S}}^+(k)
&=  \bm{T} \bigl( \mathds{1}-  i\bm{V}^{-1} \bm{\bar \gamma}^\dagger \biggl(-i \bm{\mathcal{V}}^- k' \delta   + \bm{\bar \Gamma}^{(L)}\biggr)^{-1} 
  \biggl(\mathds{1}+\bm{\bar \Gamma}^{(NL)} \bigl(-i \bm{\mathcal{V}}^- k' \delta   + \bm{\bar \Gamma}^{(L)}\bigr)^{-1}\biggr)
  (-i\bm{\bar \gamma})\bigr)\tilde{\mathbf {S}}^-(k) 
\end{aligned}
\eeq
\comtwo{This solution maintains the full linear response of a system, whilst including all nonlinear contributions to first order.}
After imposing the polar decomposition the nonlinear part of this transformation is,
\beqs
\begin{aligned}
\label{eqs:betaPerturbative}
\beta^h_{ij}
 &= \bigr[ L^{a \rightarrow s^+}  \bar \Gamma^{(Sq)} (L^{a \rightarrow s^+})^\top \bigr]_{ij}\\
  L^{a \rightarrow s^+}&:=T V^{-1} {\bar \gamma}^\dagger \bigl(-i \mathcal{V} k' \delta   + \bar \Gamma^{(L)} \bigr)^{-1}
\end{aligned}
\eeqs
In this case, the nonlinear contribution can be understood as the pump field's nonlinear contribution in the cavity modes, $\bar \Gamma^{(Sq)}$, resonantly filtered through the linear response of the system coupling the cavities to the output modes, $L^{a \rightarrow s^+}$.

In the case of the backscattering setting considered here, we find only two relevant nonzero terms contributing to any $\beta_{ij}$, those corresponding to pair generation in the forward and backward modes of the ring. If for instance, we take the indexes $s1f$ and $i1f$, (forward propagating, bus channel, signal and idler modes)
\beqs
\begin{aligned}
\label{eqs:betaPerturbativeij}
\beta_{(s1f)(i1f)}
  &=\sum_{n\in{f,b}} L^{a \rightarrow s^+}_{(s1f) (sn)}  \bar \Gamma^{(Sq)}_{(sn)(in)} L^{a \rightarrow s^+}_{(i1f) (in)}\\
\end{aligned}
\eeqs
The functions $ L^{a \rightarrow s^+}_{ij}(k,k')$ are bijective since the linear system is separable in absolute frequency. We can understand them as a resonant filtering over $k$, and a transformation of axes. The functions $|\int dk'L^{a \rightarrow s^+}_{(s1n)(in)}(k,k')|$ and $|\int dk' L^{a \rightarrow s^+}_{(i1n)(sn)}(k,k')|$, with $n=f(b)$ for the forward, \textbf{f}, (backward, \textbf{b}) directions, are plotted on the $x$ and $y$ axes of Figures~\ref{fig:allSplitOverlays}.

\section{Polar decomposition by Stimulated Data}
\label{app:SETPolar}
A straightforward method for obtaining the spontaneous JSA of a photon source in the absence of additional spatial modes, proceeds by noting that Gaussian mode transformations act on displaced states with the very same transformation as the mode operators themselves. So for some transformation,
\beq
\begin{split}
\label{eqs:GeneralGaussianM}
\mathbf M(k,k')&=\arFour{\alpha(k,k')}{\beta(k,k')}{\beta^{*}(k,k')}{\alpha^{*}(k,k')}\\
\end{split}
\eeq
and a seed coherent state (ie. displacement) is injected into mode $j$ with spectrum $d^{-}_j(k)$, the output state contains displacement in mode $i$,
\beq
\begin{split}
\label{eqs:SETDisplacements}
d_i^{+}(k)&= \int dk' \beta_{ij}(k,k') d_j^{-}(k') \, .
\end{split}
\eeq
For two-mode squeezing $\beta$ has the structure,
\beq
\begin{split}
\label{eqs:betaStructure}
\beta=\arFour{0}{\beta_{12}}{\beta_{21}}{0}
\end{split}
\eeq
which can be obtained by stimulated emission tomography. The polar decomposition (on the support of $\beta$, which is sufficient for these purposes) can be achieved by singular value decomposition of these terms by,
\beq
\begin{split}
\label{eqs:betaPolar}
\beta_{12}&=U_1\beta^D V_2^\dagger \quad , \, 
\beta_{21}=U_2\beta^D V_1^\dagger\\
\Rightarrow \beta^h&=\arFour{0}{\beta^h_{12}}{(\beta^h_{12})^\top}{0}\quad , \, 
\beta^h_{12} = U_1 \beta^D U_2^\top\\
\end{split}
\eeq
Essentially, the left singular vectors of $\beta_{12}$ and $\beta_{21}$ comprise the basis of the symmetric part, $\beta^h$, of $\beta=\beta^h V$ (with $V$ some unitary), and thus the Hermitian part, $\mathbf{M}^h$ of $\mathbf{M}=\mathbf{M}^h\mathbf{M}^u$ in the polar decomposition. The 'standard' prescription for stimulated emission tomography, then consists of following this method even though $\beta$ comprises of additional terms beyond those in Eqs.~\ref{eqs:betaStructure} which are not measured.

\section{Linear Solution in Absolute Frequency}
\label{app:LinearAbsoluteFrequency}
The linear dynamics of systems are separable in absolute frequency, which can be readily seen by transforming to,
\beqs
\begin{aligned}
\label{eqs:FourierLinearTransform}
\xi_m &= \mathcal{V}_m k +\omega_m\\
\tilde \xi_m &= V_m k +\Omega_m\\
dk &= d \xi_m \frac{\partial k}{\partial \xi_m} = d\xi_m \mathcal{V}_m^{-1}\\
\end{aligned}
\eeqs
with $\xi$ the absolute frequency. The full expression, 
\beqs
\begin{aligned}
\label{eqs:FourierLinearAbsFreq}
\int  d k'   \biggl(-i \mathcal{V} k' \delta(k-k')   + \mathcal{V} \int  \tfrac{d t}{2\pi}\e^{i (\mathcal{V} k +  \omega ) t}\bar \Gamma\e^{-i(\mathcal{V} k'+  \omega) t} \biggr)  \tilde {\mathbf a} (k')&=-i  \int   d k'   \mathcal{V} \int  \tfrac{d t}{2\pi}\e^{i (\mathcal{V} k + \omega ) t}\bar \gamma \e^{-i({V} k'+  \Omega) t} {\mathbf s}^- (k') \\
\int  d k'   \biggl(-i \mathcal{V} \bigl(\mathcal{V}^{-1}(\xi'-\omega) \bigr) \delta \bigl(\mathcal{V}^{-1}(\xi-\xi')\bigr)   + \mathcal{V} \int  \tfrac{d t}{2\pi}\e^{i \xi  t}\bar \Gamma\e^{-i\xi' t} \biggr)  \tilde {\mathbf a} (\mathcal{V}^{-1}(\xi'-\omega))&=-i  \int   d k'   \mathcal{V} \int  \tfrac{d t}{2\pi}\e^{i \xi t}\bar \gamma \e^{-i\tilde \xi' t} {\mathbf s}^- (V^{-1}(\tilde \xi'-\Omega)) \\
\end{aligned}
\eeqs
becomes separable for each frequency $\xi$ as,
\beqs
\begin{aligned}
\label{eqs:FourierLinearSol}
  \biggl(-i (\xi-\omega)   + \mathcal{V} \bar \Gamma \mathcal{V}^{-1}  \biggr)  \tilde {\mathbf a} (\mathcal{V}^{-1}(\xi-\omega))&=-i  \mathcal{V} \bar \gamma V^{-1} {\mathbf s}^- (V^{-1}( \xi-\Omega)) .
\end{aligned}
\eeqs
The input-output expression can be cast in a similar light, so that the linear system at frequency $\xi$ is described by,
\beq
\begin{split}
\label{eqs:TunedLinearFrequencyDomainInputOutputSol}
\tilde {\mathbf a} (\mathcal{V}^{-1}(\xi-\omega))&=-i \biggl(-i (\xi-\omega)   + \mathcal{V} \bar \Gamma \mathcal{V}^{-1}  \biggr)^{-1}  \mathcal{V} \bar \gamma V^{-1} {\mathbf s}^- (V^{-1}( \xi-\Omega)) \\
\tilde{\mathbf s}^+(V^{-1}( \xi-\Omega))
&= T \biggl(\mathds{1} -  V^{-1} { \bar \gamma}^\dagger\biggl(-i (\xi-\omega)   + \mathcal{V} \bar \Gamma \mathcal{V}^{-1}  \biggr)^{-1}  \mathcal{V} \bar \gamma V^{-1}  \tilde {\mathbf s}^- (V^{-1}( \xi-\Omega)) \,.
\end{split}
\eeq

\section{Nonlinear Coupling Coefficients}
\label{app:NLLambda}
The displacement field in a microring cavity supporting modes $\{\hat a_n\}_n$ is,
\beqs
\begin{aligned}
\bm{D}(\bm{r}) &= \sum_n \sqrt{\tfrac{\hbar \omega_n}{2}} \tfrac{1}{\sqrt{2 \pi}} \e^{i k(\omega_n) c}  \bm{d}^\perp_n (x,y) \hat a_n(t) + h.c.
\end{aligned}
\eeqs
where $\bm{d}^\perp_n (x,y)$ is the transverse field profile, the circumferential coordinate is $c \in(0,L]$, and the wave-vector must obey $k(\omega_n)=2\pi L/n$. Nonlinear Coupling is then
\beqs
\begin{aligned}
\hat H^{(NL)}&= -\tfrac{1}{4 \epsilon_0}\sum_{ijkl}\Gamma^{(3)}_{ijkl} \int d^3 \bm{r} \bm{D}_i(\bm{r})\bm{D}_j(\bm{r})\bm{D}_k(\bm{r})\bm{D}_l(\bm{r})\\
&=-\tfrac{1}{4 \epsilon_0}\sum_{ijkl}\tfrac{\hbar^2 \sqrt{\omega_i \omega_j \omega_k \omega_l}}{16 \pi^2} L\sinc(\Delta k L)\Gamma^{(3)}_{ijkl}\,\hat a_i(t)\hat a_j(t)\hat a_k^\dagger(t)\hat a_l^\dagger(t)\\
&\qquad \times \int dx dy \,\bm{d}^\perp_i (x,y)\bm{d}^\perp_j (x,y)\bm{d}^{\perp*}_k (x,y)\bm{d}^{\perp*}_l (x,y) + h.c.\\
&:=\hbar\sum_{ijkl} \lambda_{ijkl} \hat a_i(t)\hat a_j(t)\hat a_k^\dagger(t)\hat a_l^\dagger(t)+h.c.
\end{aligned}
\eeqs
with $\Gamma^{(3)}_{ijkl}$ defined as per reference~\cite{Sipe2004}. \com{For rings we consider, where the signal fields are closely separated from the pump, and not too narrowband we may take $\Delta k = 2\pi L (\tfrac{1}{n_i}+\tfrac{1}{n_j}-\tfrac{1}{n_k}-\tfrac{1}{n_l})\approx 0$. In the instance of very high quality factors, or significant group-velocity dispersion over the signal fields spacing, this term can become significant.}

Whilst this could couple all cavity fields, the non-negligible terms we consider are the Four-wave mixing (degenerately pumped),
\beqs
\begin{aligned}
\label{eqs:NonlinearHamiltonianFWM}
H^{(NL)FWM} &=\hbar \sum_r \lambda_{p_r p_{r}' i_r s_r} \bigl( \hat a^\dagger_{p_r} \hat a^\dagger_{p_{r}'} \hat a_{i_r} \hat a_{s_r} + \text{h.c.} \bigr)
\end{aligned}
\eeqs
where $p_r$, $p_{r}'$, $s_r$, and $i_r$ are pump, signal and idler resonances of cavity $r$, which are close to obeying the energy-matching constraint, $\omega_{p_r}+\omega_{p_{r}'}=\omega_{s_r}+\omega_{i_r}$. We also consider the Self-Phase Modulation of the pump modes,
\beqs
\begin{aligned}
H^{(NL)SPM} &=\hbar \sum_{p_r} \lambda_{p_r p_r p_r p_r} \bigl( \hat a^\dagger_{p_r} \hat a^\dagger_{p_r} \hat a_{p_r} \hat a_{p_r} + \text{h.c.} \bigr)
\end{aligned}
\eeqs
and the cross-phase modulation terms,
\beqs
\begin{aligned}
H^{(NL)XPM} &=\hbar \sum_{p_r} \sum_{n=s,i} \lambda_{p_r n_r p_r n_r } \bigl( \hat a^\dagger_{p_r} \hat a^\dagger_{n_r} \hat a_{p_r} \hat a_{n_r} + \text{h.c.} \bigr)
\end{aligned}
\eeqs

\section{Narrowband Pumping: Powers and Rates}
\label{app:pumpPower}

The displacement fields' transverse modes must obey the normalisation,
\beqs
\begin{aligned}
 1=\int dx dy \,\bm{d}^\perp_i (x,y)\bm{d}^{\perp *}_i (x,y) \frac{v_p(x,y) }{\epsilon_0 n_0(x,y)^2 v_G(x,y)}&=\int dx dy \,\bm{d}^\perp_i (x,y)\bm{d}^{\perp *}_i (x,y) \frac{c}{\epsilon_0 n_0^3(x,y) v_G(x,y)}\\
\end{aligned}
\eeqs
with $v_p(x,y)$ the local phase velocity, $v_G(x,y)$ the local group velocity, and the respective dialectric parameters are to be evaluated at the central frequency for the given effective field. We will take $n_0$ the linear refractive index in the Silicon waveguide $n_0=3.48$, and in the Silica cladding $n_0=1.46$.

In silicon, in the absence of $\chi^{(2)}$ nonlinearity, $\Gamma^{(3)}_{ijkl}$ is given by~\cite{Quesada2020},
\beqs
\begin{aligned}
\Gamma^{(3)}&=\frac{\chi^{(3)}}{\epsilon_0^{2}\left( 1+\chi^{(1)} \right)^4}\, =\frac{\chi^{(3)}}{\epsilon_0^{2}n_0^{8}}\, .
\end{aligned}
\eeqs
where $n_0$ can safely be approximated at the pump frequency. We find,
\beqs
\begin{aligned}
 \int dx dy \,\bm{d}^\perp_p (x,y)\bm{d}^\perp_p (x,y)\bm{d}^{\perp*}_s (x,y)\bm{d}^{\perp*}_i (x,y) = 3.68\times 10^{-7}\, , 
\end{aligned}
\eeqs
and using $\chi^{(3)}= 2.57\times 10^{-19} \mathrm{m^{2}V^{-2}}$ for silicon at $1.55\mathrm{\mu m}$ one obtains $\lambda=7.7\times 10^{-6} \, \text{rad}\cdot \text{s}^{-1}$.

Considering an incident coherent pulse of duration $\sigma_p=20$ps (for which a significant fraction of the input field couples to the microring, though slight additional spectral correlations result) of the form,
\beq
\av{\tilde a_p(k)}=\alpha_p \sqrt{\frac{\sigma_p {v_p}}{\sqrt{2 \pi }}} \exp \left(-\left(\frac{k \sigma_p  {v_p}}{2}\right)^2\right)
\eeq
having norm $\alpha_p$, so that it contains mean $|\alpha_p|^2$ photons. One finds that pulses of $\alpha_p=0.4\times 10^{4}$, or $2.1$pJ (equivalent to $103\mu$W average power at $50$MHz repetition rate) lead to f-f pair probability of $0.0085$ per pulse, and the generation rate at low powers is proportional to $|\alpha_p|^4$.

\section{Application: A Photonic Molecule}
\label{app:Applications}
We present one example of where the general framework could be applied, originating from the work of Chuprina et al~\cite{Chuprina2019,Chuprina2018}.

In this example, the pump channel fields $\bm s=\{a_p,f_{xp},f_{yp}\}$ and cavity modes $\bm a= \{x_p,y_p\}$ having central resonance frequencies $\bm \omega= \{\omega_{0x,p},\omega_{0y,p}\} = \{\omega_{p},\omega_{p}\}$ (where we will take $\omega_p$ from the main text) are coupled via,
\beq
\begin{aligned}
\label{eqs:ApplicationsPhotonicMoleculePump}
{\gamma} &=  \left(
\begin{array}{ccc} {\kappa_p }&{\gamma_{xp}}&{0}\\
{0}&{0}&{\gamma_{yp}} \end{array} \right), \,\, 
{g} =  \left( \begin{array}{cc} {0 }&{g_p}\\
{g_p^*}&{0} \end{array} \right)  , 
{C} =  \left(
\begin{array}{ccc} {0 }&{0 }&{0 }\\
{0}&{0 }&{0 }\\
{0 }&{0 }&{0 }\end{array} \right)\,.\\
\end{aligned}
\eeq
Here for brevity, the values in $\gamma$ will differ from those in reference~\cite{Chuprina2019} by a square root.
For the signal modes, we have channel fields $\{a_s,f_{ys},f_{zs},a_i,f_{yi},f_{zi}\}$, cavity modes $\bm a = \{z_s,y_s,z_i,y_i\}$ with central frequencies $\{\omega_{sz},\omega_{sy},\omega_{iz},\omega_{iy}\}=\{\omega_{s},\omega_{s},\omega_{i},\omega_{i}\}$ (again, $\omega_s$ and $\omega_i$ from the main text) and coupling,
\beq
\begin{aligned}
\label{eqs:ApplicationsPhotonicMoleculeSignals}
{\gamma} &=  \left(
\begin{array}{cccccc} {\kappa_s}&{\gamma_{zs}}&{0}&{0}&{0}&{0}\\
{0}&{0}&{\gamma_{ys}}&{0}&{0}&{0}\\
{0}&{0}&{0}&{\kappa_i}&{\gamma_{zi}}&{0}\\
{0}&{0}&{0}&{0}&{0}&{\gamma_{yi}} \end{array} \right), \,\, 
{g} = \left(
\begin{array}{cccc} {0}&{g_s}&{0}&{0}\\
{g_s^*}&{0}&{0}&{0}\\
{0}&{0}&{0}&{g_i}\\
{0}&{0}&{g_i^*}&{0} \end{array} \right), 
{C} =  \left(
\begin{array}{ccccccc} {0 }&{0 }&{0 }&{0 }&{0 }&{0 }\\
{0 }&{0 }&{0 }&{0 }&{0 }&{0 }\\
{0 }&{0 }&{0 }&{0 }&{0 }&{0 }\\
{0 }&{0 }&{0 }&{0 }&{0 }&{0 }\\
{0 }&{0 }&{0 }&{0 }&{0 }&{0 }\\
{0 }&{0 }&{0 }&{0 }&{0 }&{0 }\\\end{array} \right)\,.\\
\end{aligned}
\eeq
Nonlinearities occur at $\lambda_{22,24}$,$\lambda_{22,42}=\lambda$ (and we will take $\lambda$ as in the main text), ie. between the pump field $y_p$ (degenerately) and $y_s$ and $y_i$.
By appropriate choice of $\Omega$ and $V$ (taking again from the main text), and bichromatic input pump field (to be determined below), one arrives directly at the dynamics of the given system via the method here presented. 

\begin{figure}[h]
\centering
\includegraphics[width=0.8\columnwidth]{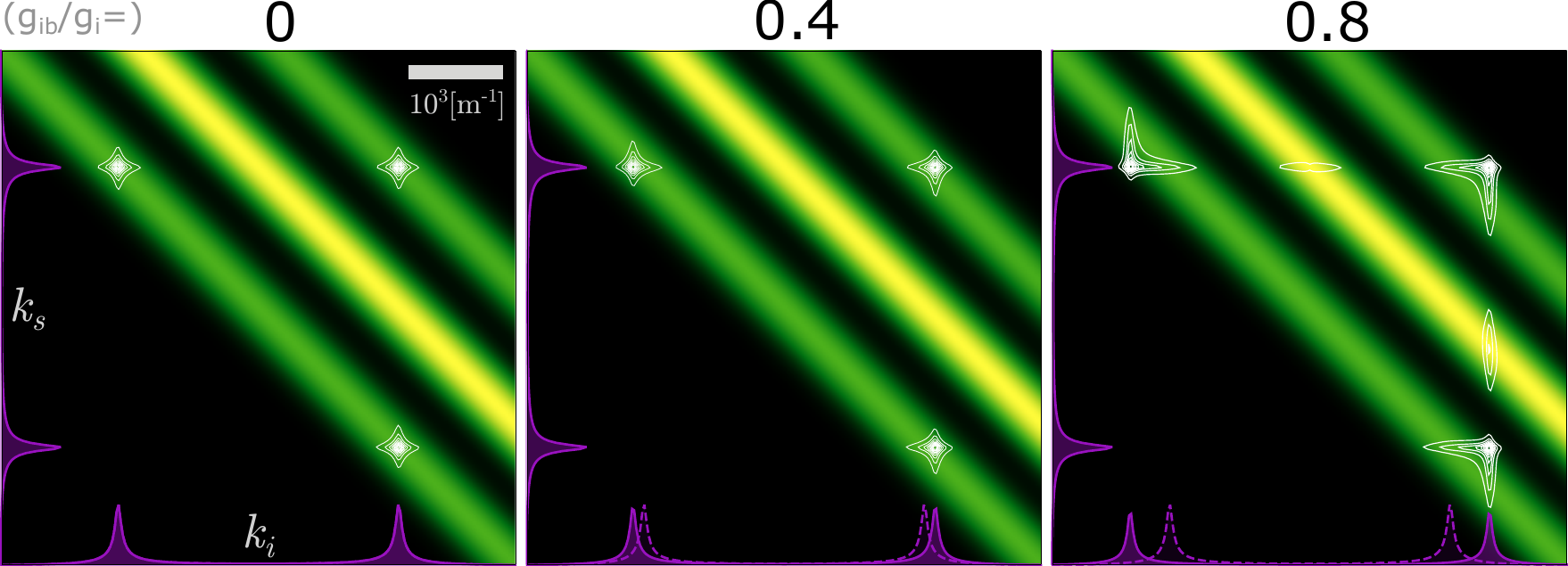}
\caption{For three values of backscatter $g_{ib}/g_i=\{0,0.4,0.8\}$, we plot the output JSA (white, contours), the nonlinear contribution ${\bar \Gamma}^{\text{Sq}}(k,k')$ (green, density), and axes plots (purple) showing the linear transmission from ring $y$ to output bus, $L^{a \rightarrow s^+}$ (Appendix.~\ref{app:pertubative}), for the given instance of $g_{ib}$ (solid, purple) as well as that with no backscatter, $g_{ib}=0$ (dashed, purple). Each is plotted over the range $k\in(-2724.6,2724.6)m^{-1}$.}
\label{fig:photonicMoleculeOverlays}
\end{figure}

Going a little further one could consider what happens if a backward propagating mode $y_{bi}$ is introduced to generation ring (ring $y$) at idler frequency along with a corresponding loss channel, $f_{ybi}$, by appending it to the signal fields,
 $\bm s=\{a_s,f_{ys},f_{zs},a_i,f_{yi},f_{zi},f_{ybi}\}$, cavity modes $\bm a = \{z_s,y_s,z_i,y_i,y_{bi}\}$ and coupling,
\beq
\begin{aligned}
\label{eqs:ApplicationsPhotonicMoleculeSignalsBackscatter}
{\gamma} &=  \left(
\begin{array}{ccccccc} {\kappa_s}&{\gamma_{zs}}&{0}&{0}&{0}&{0}&{0}\\
{0}&{0}&{\gamma_{ys}}&{0}&{0}&{0}&{0}\\
{0}&{0}&{0}&{\kappa_i}&{\gamma_{zi}}&{0}&{0}\\
{0}&{0}&{0}&{0}&{0}&{\gamma_{yi}}&{0}\\
{0}&{0}&{0}&{0}&{0}&{0}&{\gamma_{ybi}} \end{array} \right), \,\, 
{g} = \left(
\begin{array}{ccccc} {0}&{g_s}&{0}&{0}&{0}\\
{g_s^*}&{0}&{0}&{0}&{0}\\
{0}&{0}&{0}&{g_i}&{0}\\
{0}&{0}&{g_i^*}&{0}&{g_{ib}}\\
{0}&{0}&{0}&{g_{ib}^*}&{0}\end{array} \right), 
{C} =  \left(
\begin{array}{cccccccc} {0 }&{0 }&{0 }&{0 }&{0 }&{0 }&{0 }\\
{0 }&{0 }&{0 }&{0 }&{0 }&{0 }&{0 }\\
{0 }&{0 }&{0 }&{0 }&{0 }&{0 }&{0 }\\
{0 }&{0 }&{0 }&{0 }&{0 }&{0 }&{0 }\\
{0 }&{0 }&{0 }&{0 }&{0 }&{0 }&{0 }\\
{0 }&{0 }&{0 }&{0 }&{0 }&{0 }&{0 }\\
{0 }&{0 }&{0 }&{0 }&{0 }&{0 }&{0 }\\\end{array} \right)\,,\\
\end{aligned}
\eeq
where we introduced the backscattering parameter $g_{ib}$. Define the pump field,
\beq
\av{\tilde{a_p}(k)} = \frac{\alpha_p}{\sqrt{2}} \sqrt{\frac{\sigma_p {v_p}}{\sqrt{2 \pi }}}\left( \exp \left\{-\left(\frac{\left((k-\half \delta_p) +\half \delta_p\right) \sigma_p  {v_p}}{2}\right)^2\right\}+ \exp \left\{-\left(\frac{\left((k-\half \delta_p) -\half \delta_p\right) \sigma_p  {v_p}}{2}\right)^2\right) \right\} \, ,
\eeq
 setting the losses in the rings equal $\gamma_{zs}=\gamma_{ys}=\gamma_{zi}=\gamma_{yi}=\gamma_{ybi}=\gamma_{xp}=\gamma_{yp}=\gamma'=0.5 \gamma$ (ie. half the loss, $\gamma$, from the main text). We set the coupling parameters to $\kappa_s=\kappa_i=\gamma'$, and $\kappa_p=30\gamma'$, $v_s g_s=v_i g_i= 50 \gamma'^2 $, and $g_p=250 \gamma'^2 / v_p$ and the pump bandwidth $\sigma_p= 8 \sqrt{\ln(2)}/(0.6 g_s)$. Then we must choose,
\beq
\delta_{p}= \frac{ \sqrt{8 g_s^2-\frac{\gamma_{ys}^4}{v_s^2}-\left(\frac{\gamma_{zs}^2}{v_s}+\frac{\kappa_s^2}{v_s}\right)^2}}{ \sqrt{2} v_s}
\eeq
In Fig~\ref{fig:photonicMoleculeOverlays} we plot the JSA and the perturbative contributions resulting from idler backscattering $g_{ib}/g_{i}=\{0,0.4,0.8\}$. 

\section{Derivation of Frequency Domain Expressions}
\label{app:FourierInduction}
The time domain equations of motion are linear and thus admit linear solutions in the Fourier domain. We move to the frequency domain by applying the Fourier transform, and substituting the operators' for their Fourier components. For the linear equations of motion (Eqs.~\ref{eqs:linearEOMsTime}) we arrive at Eqs.~\ref{eqs:FourierLinear} via,
\beqs
\begin{split}
\label{eqs:FourierLinearInduction}
\mathcal{V} \int  \tfrac{d t}{\sqrt{2\pi}} \e^{i \mathcal{V} k t} \biggl(\frac{d}{dt}  +\bar \Gamma(t) \biggr) \tilde {\bm a} (t)&=\mathcal{V} \int  \tfrac{d t}{\sqrt{2\pi}} \e^{i \mathcal{V} k t} \bigl(- i \bar \gamma(t)  \tilde {\bm s}^- (0,t)\bigr)\\
\mathcal{V} \int  \tfrac{d t}{\sqrt{2\pi}} \e^{i \mathcal{V} k t} \biggl(-i \mathcal{V} k'  +\bar \Gamma(t) \biggr) \int  \tfrac{d k'}{\sqrt{2\pi}}\e^{-i \mathcal{V} k' t} \tilde {\bm a} (k')&=-i \mathcal{V} \int  \tfrac{d t}{\sqrt{2\pi}} \e^{i \mathcal{V} k t} \bar \gamma(t)  \int \tfrac{d k'}{\sqrt{2\pi}} \e^{-iV k' t}  {\bm s}^- (k') \\
\int  d k'   \biggl(-i \mathcal{V} k' \delta(k-k')   + \mathcal{V} \int  \tfrac{d t}{2\pi}\e^{i \mathcal{V} k t}\bar \Gamma(t)\e^{-i \mathcal{V} k' t} \biggr)  \tilde {\bm a} (k')&=-i  \int   d k' \biggl(\mathcal{V}\int \tfrac{d t}{2\pi}\e^{i \mathcal{V} k t} \bar \gamma(t)    \e^{-iV k' t}\biggr)  {\bm s}^- (k') \\
\int  d k'   \biggl(-i \mathcal{V} k' \delta(k-k')   + \bar \Gamma(k,k') \biggr)  \tilde {\mathbf a} (k')
&=-i  \int   d k'  \bar \gamma(k,k')      {\mathbf s}^- (k') \,.
\end{split}
\eeqs

For the nonlinear equations of motion, some care must be taken to ensure the Fourier transform is applied consistently. Given Eqs.~\ref{eqs:nonlinearEOMsTime} we arrive at Eqs.~\ref{eqs:FourierNonLinear} via,
\beqs
\begin{split}
\label{eqs:FourierNonLinearInduction}
\bm{\mathcal{V}} \int  \tfrac{d t}{\sqrt{2\pi}} \e^{i \bm{\mathcal{V}}^- k t} \biggl(\frac{d}{dt}  +\bm{\bar \Gamma}(t) \biggr) \tilde {\bm A} (t)&=\bm{\mathcal{V}} \int  \tfrac{d t}{\sqrt{2\pi}} \e^{i  \bm{\mathcal{V}}^- k t} \bigl(- i \bm{\bar \gamma}(t)  \tilde {\bm S}^- (0,t)\bigr)\\
\bm{\mathcal{V}} \int  \tfrac{d t}{\sqrt{2\pi}} \e^{i \bm{\mathcal{V}}^- k t} \biggl(-i \bm{\mathcal{V}}^- k'  +\bm{\bar \Gamma}(t) \biggr) \int  \tfrac{d k'}{\sqrt{2\pi}}\e^{-i \bm{\mathcal{V}}^- k' t} \tilde {\bm A} (k')&=-i \bm{\mathcal{V}} \int  \tfrac{d t}{\sqrt{2\pi}} \e^{i \bm{\mathcal{V}}^- k t} \bm{\bar \gamma}(t)  \int \tfrac{d k'}{\sqrt{2\pi}} \e^{-i \bm{V}^- k' t}  {\bm S}^- (k') \\
\int  d k'   \biggl(-i \bm{\mathcal{V}}^- k' \delta(k-k')   + \bm{\mathcal{V}} \int  \tfrac{d t}{2\pi}\e^{i \bm{\mathcal{V}}^- k t}\bm{\bar \Gamma}(t)\e^{-i \bm{\mathcal{V}}^- k' t} \biggr)  \tilde {\bm A} (k')&=-i  \int   d k' \biggl(\bm{\mathcal{V}}\int \tfrac{d t}{2\pi}\e^{i \bm{\mathcal{V}}^- k t} \bm{\bar \gamma}(t)    \e^{-i \bm{V}^- k' t}\biggr)  {\bm S}^- (k') \\
\int  d k'   \biggl(-i \bm{\mathcal{V}}^- k' \delta(k-k')   + \bm{\bar \Gamma}(k,k') \biggr)  \tilde {\mathbf A} (k')\nonumber
&=-i  \int   d k'  \bm{\bar \gamma}(k,k')      {\mathbf S}^- (k'),
\end{split}
\eeqs
for which we indeed find the linear solution.

\end{appendices}

\addcontentsline{toc}{chapter}{Bibliography}

\bibliography{library}
\bibliographystyle{unsrt}

\end{document}